\documentclass[twocolumn,prb,showpacs,floatfix]{revtex4}
\usepackage{amsmath}
\usepackage{amssymb}
\usepackage{bm}
\usepackage{graphicx}
\usepackage{verbatim}
\usepackage{color}
\usepackage{subfigure}

\begin{document}

\title{Nuclear Dynamics During Landau-Zener Singlet-Triplet Transitions in
Double Quantum Dots}

\author{Arne Brataas$^{1,2}$ and Emmanuel I. Rashba$^{1}$}
\affiliation{Department of Physics, Harvard University, Cambridge, Massachusetts 02138,
USA\\
Department of Physics, Norwegian University of Science and Technology,
NO-7491 Trondheim, Norway}

\begin{abstract}
We consider nuclear spin dynamics in a two-electron double dot system near
the intersection of the electron spin singlet $S$ and the lower energy
component $T_{+}$ of the spin triplet. The electron spin interacts with
nuclear spins and is influenced by the spin-orbit coupling. Our approach is
based on a quantum description of the electron spin in combination with the
coherent semiclassical dynamics of nuclear spins. We consider single and
double Landau-Zener passages across the $S$-$T_{+}$ anticrossings. For
linear sweeps, the electron dynamics is expressed in terms of parabolic
cylinder functions. The dynamical nuclear polarization is described by two
complex conjugate functions $\Lambda ^{\pm }$ related to the integrals of
the products of the singlet and triplet amplitudes ${\tilde{c}}_{S}^{\ast }{%
\tilde{c}}_{T_{+}}$ along the sweep. The real part $P$ of $\Lambda ^{\pm }$
is related to the $S$-$T_{+}$ spin-transition probability, accumulates in
the vicinity of the anticrossing, and for long linear passages coincides
with the Landau-Zener probability $P_{LZ}=1-e^{-2\pi \gamma }$, where $%
\gamma $ is the Landau-Zener parameter. The imaginary part $Q$ of $\Lambda
^{+}$ is specific for the nuclear spin dynamics, accumulates during the
whole sweep, and for $\gamma \gtrsim 1$ is typically an order of magnitude
larger than $P$. $P$ and $Q$ also show critically different dependences on
the shape and the duration of the sweep. $Q$ has a profound effect on the
nuclear spin dynamics, by (i) causing intensive shake-up processes among the
nuclear spins and (ii) producing a high nuclear spin generation rate when
the hyperfine and spin-orbit interactions are comparable in magnitude. Even
in the absence of spin-orbit coupling, when the change in the the total
angular momentum of nuclear spins is less than $\hbar $ per single
Landau-Zener passage, the change in the global nuclear configuration might
be considerably larger due to the nuclear spin shake-ups. We find analytical
expressions for the back-action of the nuclear reservoir represented via the
change in the Overhauser fields the electron subsystem experiences.
\end{abstract}

\maketitle
\date{\today }



\section{Introduction}

\label{sec:intro}


Electron spin states in semiconductor quantum dots are investigated for
their potential use as quantum bits in quantum computing architectures.\cite%
{Hanson:rmp07,Loss:prb98,Levy:prl02} To this end, control of the spin states
and their couplings to the environment is essential. In GaAs and InAs
semiconductors, a major source of electron spin decoherence is the coupling
to the surrounding nuclear spins.\cite%
{Merkulov:prb02,Khaetskii:prb03,Erlingsson:prb04,Hanson:rmp07,LuSham,Cywinski}
Since the quantum dots are large compared to the interatomic spacing, each
electron interacts with typically one million nuclei. Achieving control over
this many-body interaction is a key for manipulating semiconductor quantum
bits.

In two electron double quantum dots, the singlet $S$ and triplet $T_0$
states define the elementary qubit. The coupling between these states is
governed by the gradient in the longitudinal magnetic Zeeman splitting
between the two dots. Controlling this coupling enables singlet-triplet
qubit manipulations. Beyond the two-state $S$-$T_0$ qubit operation, the
gradient in the transverse magnetic Zeeman splitting between the two dots
defines the coupling of the singlet $S$ to the triplet $T_+$ and $T_-$
states. Finally, the longitudinal magnetic Zeeman splitting determines the
relative energies of the triplet states. This Zeeman splitting arises from
the external field $\mathbf{B}$ and the nuclear spin background via the
Overhauser field, and by changing the nuclear spin polarization the basic
electron parameters can be tuned.

Polarization of nuclear spins can be created and destroyed by flip-flop
processes by pumping the electronic states via time-dependent gate voltages.
This has recently been investigated in many interesting experimental\cite%
{Barthel,Foletti:nphys09,Bluhm:prl10,Bluhm:nphys11} and theoretical papers
in double quantum dots in the regime of Pauli blockade.\cite%
{Ramon:prb07,Gullans:prl10,Stopa:prb10,Rudner:prb10a,Rudner:prb10b}
Experimentally, it has been demonstrated that an Overhauser field gradient
of several hundred milli Tesla can be generated and sustained.\cite%
{Foletti:nphys09} The dephasing time of the electron-spin qubits has been
extended to more than 200 $\mu s$.\cite{Bluhm:nphys11} Because the dynamical
interaction of an electron spin with a nuclear spin reservoir is enormously
complicated, different theoretical efforts were focused on the various
aspects of it. The two aspects most closely related to our paper are the
theoretical modeling of the connection between the generation of dynamical
nuclear spin polarization at short and long time scales\cite%
{Ramon:prb07,Gullans:prl10,Burkard} and the influence of the spin-orbit
interaction on the build-up of the nuclear polarization.\cite%
{Rudner:prb10a,Rudner:prb10b}

The aim of this paper is to study in detail the electron and nuclear spin
dynamics as the system passes across a $S$-$T_{+}$ anticrossing. In GaAs and
InAs quantum dots in an external magnetic field, $T_{+}$ is the lowest
energy component of the electron triplet state because of the negative
electron $g$-factor, $g<0$. During a $S$-$T_{+}$ (or a $T_{+}$-$S$) passage,
electrons trade their spin with the nuclear reservoir, and multiple passages
are used in creating a difference ("gradient") of the effective nuclear
(Overhauser) fields between two parts of the double dot that are used for
qubit rotations. The study of a single passage (or two passages during a
single cycle) provides a firm basis for investigating events on longer time
scales. Also, the progress in experimental techniques currently allows,
instead of averaging data over thousands of sweeps, to perform single-shot
measurements,\cite{SingleShot} and most recently such measurements have been
achieved for double quantum dots.\cite{Barthel} Also, the double dot
dynamics during a single sweep manifests itself explicitly in beam splitter
experiments.\cite{Petta2010} We expect the approach developed in our paper
to become a useful tool in discussing such types of experiments and, more
widely, to facilitate better understanding and utilization of the nuclear
spin environment in solid state based quantum computing.

Specifically, we take into account the spatial distribution of the hyperfine
coupling between the electron and nuclear spins and compute the change in
the topography of the nuclear spin polarization and the related changes in
the gradient and average Overhauser fields governing the dynamics of the
electron spin. These fields, that the electrons experience in the singlet
and triplet states, depend on the spatial variation of the electron-nuclear
coupling and we take this dependence into account. We employ the Zener
approach\cite{Zener:prs32} and find analytically explicit expressions for
the electron and nuclear spin dynamics during a single linear sweep and
during cycles consisting of two linear sweeps.

Let us give an overview of the main results. We express the whole electron
and nuclear spin dynamics in terms of two complex conjugate functions $%
\Lambda ^{\pm }(T_{i},T_{f})$ depending on the initial and finite times $%
(T_{i},T_{f})$ and the shape of the path between them. These $\Lambda^\pm$
functions are integrals of the products of the singlet and triplet
amplitudes during the $S$-$T_+$ passage. The real part $P=\text{Re}
\{ \Lambda^\pm \}$ is the transition probability between the singlet $S$ and
the triplet $T_+$ states. The imaginary part $Q=\text{Im} \{ \Lambda^+\}$
includes basic information about the nuclear spin dynamics including the
nuclear shake-ups. The Landau-Zener probability, $P_{LZ}=1-e^{-2\pi \gamma }$%
, where $\gamma $ is the Landau-Zener parameter, is the asymptotic value of $%
P(T_i,T_f)$ for a single sweep when $T_{i}\rightarrow -\infty$ and $%
T_{f}\rightarrow \infty $. Usually, all results are expressed in terms of $%
P_{LZ}$. Our approach provides a more detailed information about the nuclear
spin dynamics away from the $S-T_{+}$ anticrossing.

Oscillations of the transition probability $P(T_i,T_f)$ as a function of its
arguments reveal typical interference patterns. These oscillations are
highly anharmonic for small Landau-Zener transition probabilities $P_{LZ}\ll
1$ and might persist for a long time with a large amplitude for intermediate
Landau-Zener transition probabilities $P_{LZ}\sim 0.5$. However, it is not
typically the transition probability $P$ that determines the nuclear spin
dynamics. Instead, the other $S$-$T_+$ quantity, $Q$ is non less important.
While $P$ is constrained to be in the interval $0\leq P\leq1$, there are no
such constraints on $Q$ and it is typically larger than $P$. We find that $Q$
controls the shake-up processes among the nuclear spins. In the absence of
spin-orbit coupling, at most $\hbar$ of the angular momentum can be
transferred to the nuclear spin bath. Given that there are around a million
nuclear spins in the quantum dots, of which around a thousand are aligned
initially, a change in one out of a thousand nuclear spins would have only a
minor effect. However, the nuclear spins are allowed to interchange their
spins during the $S$-$T_+$ passage without violating the conservation of the
angular momentum. Although the interchange does not change the total nuclear
spin angular momentum, the redistribution of the nuclear spins can lead to
considerable changes in the various gradient and average Overhauser fields
that the electrons experience. This is because the Overhauser fields depend
on weighted average values of the nuclear spin distribution with respect to
the electron-nuclear couplings and not just the total nuclear spin. We find
that such shake-ups are very sensitive to the initial nuclear spin
distribution and that they are often much larger than the average nuclear
spin production because $Q$ is typically ten times larger than $P$.

Furthermore, when the spin-orbit coupling competes with the hyperfine
interaction and $Q$ is considerably larger than $P$, then the $Q$-enhanced
spin generation dominates for a generic direction of the nuclear spin
polarization and can become considerably larger than $P$. However, after
averaging over the direction of the transverse nuclear spin polarization, $Q$
cancels and the results of Refs. [\onlinecite{Rudner:prb10a,Rudner:prb10b}]
are recovered.

Another finding is that even geometrically symmetric double quantum dots
acquire asymmetric behavior because of the spatial inhomogeneity of the
hyperfine coupling. The sign of the asymmetry depends on $B$, and its
magnitude is largest close to the (0,2) or (2,0) configuration. The
consequences of this $B$-controlled asymmetry for building nuclear field
gradients are similar to that envisioned in Ref.~\onlinecite{Gullans:prl10}
for geometrically asymmetric dots.

This paper is organized in the following way. In Sec. \ref{sec:model}, we
describe the model of a double quantum dot that follows the lines of Refs. [%
\onlinecite{Taylor:nphys05,Taylor:prb07,Gullans:prl10}]. We introduce the
basic notations related to the electron-nuclear hyperfine interaction and
the nuclear dynamics induced by it in Secs.~\ref{sec:electronnucleardynamics}
and \ref{sec:dynamicalnuclearpolarization}, respectively. In Sec. \ref%
{sec:LandauZener}, a linear Landau-Zener sweep is treated analytically and
the time-dependence of the effective magnetic fields acting on the nuclei is
discussed in detail. Because Sec. \ref{sec:LandauZener} is rather technical,
a reader interested in experimental applications can skip to Sec. \ref%
{sec:STS}, where numerical data for the linear in time Landau-Zener sweeps
and cycles are discussed. In Sec. \ref{sec:SUP}, the back action of the
nuclear spin dynamics on the Overhauser fields in the electron spin
Hamiltonian is estimated. Appendix \ref{sec:spinoperator} outlines the
notations for electron spin operators. Appendix \ref{sec:spatial} discusses
the spatial dependence of the hyperfine interaction. We demonstrate that
even for two symmetric quantum dots, the hyperfine coupling acquires
asymmetries controlled by the overlap integral and the external magnetic
field. Appendix \ref{sec:identities} includes two new identities for
parabolic cylinder functions. We conclude and summarize our results in Sec. %
\ref{sec:conclusions}.

\section{Model}

\label{sec:model} 

We consider two electrons in a double quantum dot. When the electron spin is
conserved, the classification of the electron states as a singlet state $S$
and three triplet ($T_{\nu }$, $\nu =0,\pm 1$) states is exact. Spin-orbit
interaction and the interaction with the nuclear spins mixes these states.
We use the singlet and triplet stationary states as our basis. They are 
\begin{subequations}
\label{SinTr}
\begin{align}
\Psi _{S}(1,2)& =\psi _{S}(1,2)\chi _{S}(1,2)\,,  \label{singlettripletbasis}
\\
\Psi _{T_{\nu }}(1,2)& =\psi _{T}(1,2)\chi _{T_{\nu }}(1,2)\,,
\end{align}%
%
where 1 and 2 denote the 1st and 2nd electron. The spin wave functions obey
the symmetries $\chi _{S}(1,2)=-\chi _{S}(2,1)$ as well as $\chi _{T_{\nu
}}(1,2)=\chi _{T_{\nu }}(2,1)$ and are specified in Appendix \ref%
{sec:spatial}. The orbital wave functions $\psi _{S}(1,2)$ and $\psi
_{T}(1,2)$ obey the symmetries $\psi _{S}(1,2)=\psi _{S}(2,1)$ and$\, \, \psi
_{T}(1,2)=-\psi _{T}(2,1),$ and we consider only the lowest energy orbital
states so there are no additional quantum numbers labeling the orbital wave
functions.

The electrons interact with each other, external gate potentials, an
external magnetic field, and with the nuclear spins predominantly via the
hyperfine interaction. The latter interaction, as well as spin-orbit
coupling, induce transitions between the singlet and triplet states that we
compute. The nuclei interact with the external magnetic field, the electrons
through the hyperfine interaction, and with each other via the magnetic
dipole-dipole interaction. The latter interaction affects the nuclear spin
dynamics on long time scales of around milli seconds, and we disregard it in
what follows. However, we take into account (in a semiclassical
Born-Oppenheimer approach and in the leading order in the large electron
Zeeman splitting) an indirect RKKY-like interaction between nuclear spins
originating from the hyperfine electron-nuclear coupling (see Sec.\  \ref%
{sec:QS}). Near the $ST_+$ anticrossing it manifests itself at the scale of
about 10 $\mu$s.

Of central importance is the hyperfine electron-nuclear interaction 
\end{subequations}
\begin{equation}
\hat{H}_{hf}=A\sum_{j}\sum_{\ell =1}^{2}\delta (\mathbf{{R}_{j}-r_{\ell })(%
\hat{I}_{j}\cdot \hat{\mathbf{s}}(\ell )),}  \label{Hhf}
\end{equation}
%
where $A$ is the electron-nuclear interaction strength, $\ell $ numerates
electrons and $j$ nuclei, $\hat{\mathbf{s}}(\ell)=\frac{1}{2}\hat{%
\boldsymbol{\sigma }}(\ell)$ are the electron spin operators in terms of the
vector of Pauli matrices $\hat{\boldsymbol{\sigma }}(\ell)$ for each
electron $\ell $, and $\hat{\mathbf{I_{j}}}$ are the nuclear spin operators.
The electron and nuclear spin operators are dimensionless in our notations.
Carets denote quantum mechanical operators and bold variables are vectors.

In the $4\times 4$ singlet and triplet space ($S,T_{+},T_{0}$, and $T_{-}$),
the Hamiltonian that describes the electrons and their interaction with the
nuclear spins can be written as 
\begin{equation}
{\hat{H}}=\left( 
\begin{array}{cc}
\epsilon _{S} & \mathbf{\hat{v}}_{n}^{T} \\ 
\mathbf{\hat{v}}_{n}^{\ast } & \epsilon _{T}-\mathbf{\hat{\eta}\cdot \hat{S}}%
\end{array}%
\right) ,  \label{Hfull}
\end{equation}%
%
where the total electron spin ${\hat{\mathbf{S}}}=\hat{\mathbf{s}}(1)+\hat{%
\mathbf{s}}(2)$. Additionally, the spin-orbit interaction induces terms in
Eq.~(\ref{Hfull}) that we discuss below. The nuclear spins are also affected
by the external magnetic field through the nuclear Zeeman effect that we
take into account below in the description of their dynamics. However, we
disregard the effect of the nuclear Zeeman energy on the equilibrium spin
populations because of the high temperature of the nuclear spin bath. The $%
\epsilon _{S}$ and $\epsilon _{T}$ terms in the diagonal matrix elements of
Eq. (\ref{Hfull}) describe the singlet and triplet energies in the absence
of the nuclear and external magnetic fields. They depend on the
electrostatic gate potentials and the interactions between the electrons.
The off-diagonal operator components $\mathbf{\hat{v}}_{n}^{T}=(\hat{v}%
_{n}^{+},-\hat{v}_{n}^{z},-\hat{v}_{n}^{-}),$ are nuclear spin dependent (a
superscript $T$ denotes the transpose of a vector and the subscript $n$
denotes that this coupling is due to the nuclear spins) 
\begin{equation}
\hat{v}_{n}^{\alpha }=A\sum_{j}\rho _{j}\hat{I}_{j}^{\alpha },  \label{vn}
\end{equation}%
%
with $\alpha =(+$, $-$, $z)$, $\hat{I}_{j}^{\pm }=\left( \hat{I}_{j}^{x}\pm i%
\hat{I}_{j}^{y}\right) /\sqrt{2}$ are the transverse nuclear spin
components, and the singlet-triplet electron-nuclear coupling coefficients 
\begin{equation}
\rho _{j}=\rho (\mathbf{R}_{j})=\int d\mathbf{r}\psi _{S}^{\ast }(\mathbf{r},%
\mathbf{R}_{j})\psi _{T}(\mathbf{r},\mathbf{R}_{j})
\label{singlettripletcoupling}
\end{equation}%
%
dependent on the positions $\mathbf{R}_{j}$ of nuclei $j$. Roughly, $\rho
_{j}$ varies from positive in one quantum dot to negative in the other.
Therefore, $\hat{v}_{n}^{\pm }$ and $\hat{v}_{n}^{z}$ represent differences
in the effective nuclear magnetic fields in the two dots in the directions
transverse and parallel to the external magnetic field, respectively. The
effective splitting of the triplet states due to the external magnetic field 
$\mathbf{B}$ and the nuclei is $-\mathbf{\hat{\eta}\cdot \hat{S}}$, where 
\begin{equation}
\mathbf{\hat{\eta}}=\eta _{Z}\mathbf{e}_{z}\,+\mathbf{\hat{\eta}}_{n}=\eta
_{Z}\mathbf{e}_{z}\,-A\sum_{j}\zeta _{j}\mathbf{\hat{I}}_{j}\,,
\label{etaHat}
\end{equation}%
%
$\eta _{Z}$ is the electron Zeeman splitting in the 
field $\mathbf{B}\parallel {\hat{\mathbf{z}}}$, $\mathbf{\hat{S}}$ is the
spin-1 operator for the electrons (as defined in Appendix \ref%
{sec:spinoperator}), and the position dependent coupling constants of the
triplet states to the nuclei are 
\begin{equation}
\zeta _{j}=\int d\mathbf{r}\psi _{T}^{\ast }(\mathbf{r},\mathbf{R}_{j})\psi
_{T}(\mathbf{r},\mathbf{R}_{j}).  \label{triplettripletcoupling}
\end{equation}%
%
This completes the description of the Hamiltonian that governs the coupling
between the electron 
and nuclear spin dynamics.

The $ST_{+}$ anticrossings arising due to $\hat{v}_{n}^{\alpha }$ and also
the $ST_{0}$ level splittings were investigated by the beam-splitting
technique\cite{Petta2010} and Rabi-oscillations \cite%
{Petta2005,Foletti:nphys09,Bluhm:prl10}, respectively.

\section{Electron and nuclear spin dynamics}

\label{sec:electronnucleardynamics}


The Hamiltonian of Eq.~(\ref{Hfull}) defines a many-body problem of the
coupled electron-nuclear dynamics. Our interest is in the dynamical nuclear
polarization that is achieved by changing the gate voltages in such a way
that the electronic subsystem makes a transition from the singlet $S$ to the
lowest energy triplet $T_{+}$ state or vice versa. The many-body interaction
can be simplified by employing the Born-Oppenheimer approach.\cite%
{Gullans:prl10} The electrons are fast as compared to the nuclei. The
electrons also interact with a large number of nuclei, around one million.
These two features imply that the electron dynamics is unaffected by the
dynamics of a single nucleus and electrons see only a quasi-static
configuration of all nuclei during a single $ST_{+}$ crossing. This
motivates an ansatz where the wave function is separable into electronic and
nuclei parts.\cite{Gullans:prl10}

The electron dynamics can be solved from the Hamiltonian of Eq.~(\ref{Hfull}%
) with the assumption that the nuclear spin operators can be approximated by
their expectation values before the transition, $\mathbf{\hat{v}}%
_{n}\rightarrow \mathbf{v}_{n}$. The detuning energy $\epsilon $ is defined
as the difference between the triplet energy $\epsilon _{T_{0}}$ and the
singlet energy $\epsilon _{S}$, $\epsilon =\epsilon _{T_{0}}-\epsilon _{S}$,
and is controlled by the variations in the gate voltages. We restrict
ourselves to the limit of a rather large external magnetic field so that the
splitting between the triplet states is larger than the magnitude of the
off-diagonal matrix elements that mix the singlet and triplet states. When
the separation between the energy levels is much larger than the matrix
elements that mix the singlet and triplet states, the singlet and triplet
states are well separated. The singlet-triplet matrix elements produce
anti-crossings between the singlet and triplet levels when their energies
are tuned to be close to resonance. Our focus is on situations where the
system is tuned close to the $S$-$T_{+}$ transition as shown in Fig. \ref%
{fig:st+} 
\begin{figure}[tbph]
\includegraphics[width=0.9\columnwidth]{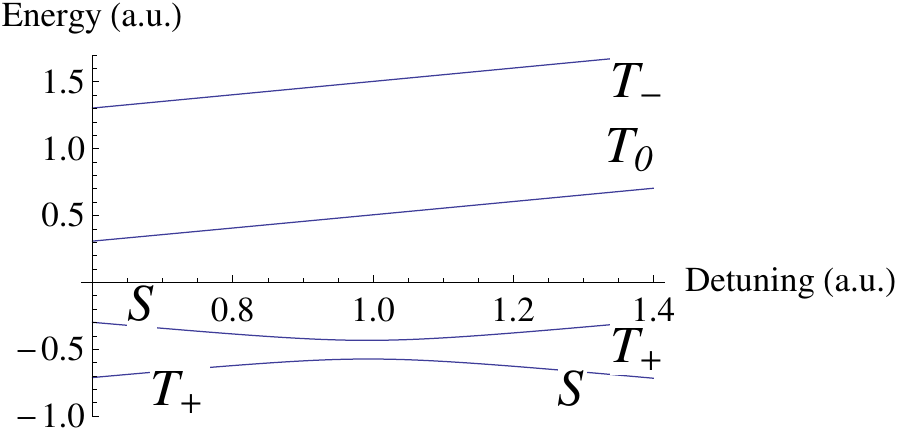}
\caption{Schematics of the singlet and triplet energy levels as a function
of the detuning energy $\protect \epsilon =\protect \epsilon _{T_{0}}-\protect%
\epsilon _{S}$ close to the $S$-$T_{+}$ anticrossing. The Zeeman splitting $%
\protect \eta _{Z}=1$ is chosen as the energy unit, off-diagonal matrix
elements are $v^{\perp }=|v^{\pm }|=0.07$.}
\label{fig:st+}
\end{figure}
There, the energies of the triplet states $T_{0}$ and $T_{-}$ are of the
order the electron Zeeman splitting $\eta ^{z}$ away from the energies of
singlet $S$ and triplet $T_{+}$ states, which is a large energy as compared
to the $S$-$T_{+}$ anticrossing width. In this case, the electron dynamics
can be approximated by the $2\times 2$ dynamics for the singlet $S$ and
triplet $T_{+}$ amplitudes of the electron wave function. The reduction of
the original $4\times 4$ electron dynamics problem to a $2\times 2$ problem
also facilitates finding an exact solution for the electron dynamics for
linear sweeps and allows to reveal the role of the long time
\textquotedblleft tails" of the singlet and triplet amplitudes crucial for
the nuclear spin dynamics. In the $2\times 2$ basis, the electron dynamics
is described by the singlet $c_{S}$ and triplet $c_{T_{+}}$ amplitudes that
obey a Schr\"{o}dinger equation 
\begin{equation}
H^{(ST_{+})}\left( 
\begin{array}{c}
c_{S} \\ 
c_{T_{+}}%
\end{array}%
\right) =i\hbar \partial _{t}\left( 
\begin{array}{c}
c_{S} \\ 
c_{T_{+}}%
\end{array}%
\right)   \label{DynEq}
\end{equation}%
%
with the Hamiltonian 
\begin{equation}
H^{(ST_{+})}=\left( 
\begin{array}{cc}
\epsilon _{S} & v^{+} \\ 
v^{-} & \epsilon _{T_{+}}%
\end{array}%
\right) ,  \label{Hst+}
\end{equation}%
%
where $\epsilon _{T_{+}}=\epsilon _{T}-\eta ^{z}$, and following Refs. [%
\onlinecite{Rudner:prb10a,Rudner:prb10b}] we have included the spin-orbit
matrix elements $v_{so}^{\pm }$ that couple $S$ and $T_{+}$ states into the
total off-diagonal matrix elements 
\begin{equation}
v^{\pm }=v_{n}^{\pm }+v_{so}^{\pm }.  \label{vtotal}
\end{equation}%
%
While the coupling between $S$ and $T$ levels in GaAs double quantum dots is
usually attributed to the hyperfine interaction, spin-orbit coupling is
inevitably present while difficult to evaluate quantitively for specific
devices.\cite{Novak2011} It manifests itself in spin relaxation,\cite%
{KhaetskiiNazarov,BulaevLoss} level anticrossings in InAs single and double
dots,\cite{Takahashi,Pfund2007} and in the EDSR\cite{EDSR,Golovach} both in
GaAs\cite{Nowack07,Tarucha08} and InAs\cite{NadjPerge} double dots. It is
important to emphasize the existence of different mechanisms that couple the
electron spin to the orbital degrees of freedom. They include the
traditional (Thomas) spin-orbit interaction that couples the electron spin
to the electron momentum and the Zeeman interaction in a inhomogeneous
magnetic field $\mathbf{B}(\mathbf{r})$ that couples the electron spin to
the elecron coordinate.\cite{Rashba05} In Ref.~\onlinecite{Nowack07}, the
first mechanism dominated while in Refs.~\onlinecite{Laird07} and %
\onlinecite{Tarucha08} different versions of the second one were important.
We show in what follows that spin-orbit coupling also has a profound effect
on the nuclear spin polarization production rate.

By carrying out a unitary transformation of the original $4\times 4$
Hamiltonian, it can be shown that the corrections to the reduced $2\times 2$
Hamiltonian of Eq.~(\ref{Hst+}) are quadratic in the small ratio between $%
v^{\pm }$ and the Zeeman splitting $\eta_{Z}$ provided the gate-voltage
induced $S$-$T_{+}$ transition is slow so that $\hbar (\dot{\epsilon}_{S}-%
\dot{\epsilon}_{T})/\eta_{Z}^{2}\ll 1$. We assume that this criterion is
satisfied.

In turn, the dynamics of nuclear spins is driven by the effective magnetic
fields $\boldsymbol{\Delta }_{j}$ arising from the electron dynamics 
\begin{equation}
\hbar \frac{d\mathbf{\hat{I}}_{j}}{dt}=\boldsymbol{\Delta }_{j}\times 
\mathbf{\hat{I}}_{j},  \label{NucDyn}
\end{equation}
%
where the components of the fields $\boldsymbol{\Delta }_{j}$ acting on the
nuclei are the transverse $\Delta _{j}^{\pm }=\left( \Delta _{j}^{x}\pm
i\Delta _{j}^{y}\right) /\sqrt{2}$ and longitudinal $\Delta _{j}^{z}$
fields: 
\begin{subequations}
\begin{align}
\Delta _{j}^{+}& =A\rho _{j}c_{S}c_{T_{+}}^{\ast },  \label{Delta+} \\
\Delta _{j}^{-}& =A\rho _{j}c_{S}^{\ast }c_{T_{+}},  \label{Delta-} \\
\Delta _{j}^{z}& =A\zeta _{j}|c_{T_{+}}|^{2}-\eta _{j(nZ)},  \label{Deltaz}
\end{align}
%
and $\eta _{j(nZ)}$ is the nuclear Zeeman splitting for the nucleus $j$.
Because the dynamics of electron amplitudes $(c_S(t),c_{T_+}(t))$ depends
not only on the potentials on the gates but also on the nuclear spins
through the matrix elements $v^\pm$, fields $\boldsymbol{\Delta }_{j}$ can
be considered as dynamical RKKY fields.

In the next section we show how the changes in the electronic states as they
pass across the $S$-$T_{+}$ anticrossing change the spatially dependent
nuclear polarization.

\section{Dynamical Nuclear Polarization}
\label{sec:dynamicalnuclearpolarization}


We consider a situation where the changes in the gate voltages can induce a
singlet $S$ to triplet $T_{+}$ transition or vice versa, so that the total
electron angular momentum may be increased or reduced by 1. In the absence
of spin-orbit coupling, this implies that the change in the $z$-projection
of the total nuclear spin equals the change in the elecron spin (but with
the opposite sign). There is no conservation law for the spatial
distribution of the nuclear spin. We are interested in how this change of
angular momentum is distributed among the nuclei. As already mentioned
above, the typical time scale for  nuclei dynamics is 
long as compared to the time scale for the electron dynamics, in particular,
with the singlet-triplet transition time. Let us denote the initial time of
the sweep as $T_{i}$ and the final time as $T_{f}$. We assume that the
duration of the Landau-Zener sweep, $T_{f}-T_{i}$, is short as compared to
the typical nuclear spin precession time and take the nuclear dynamics into
account as a perturbation. Also, since the total change of the angular
momentum is of the order $1$, the typical change in the individual nuclear
spins is much less than $1$. With these assumptions, the change of a nuclear
spin $\Delta \mathbf{\hat{I}}_{j}=\mathbf{\hat{I}}_{j}(T_{f})-\mathbf{\hat{I}%
}_{j}(T_{i})$ during a Landau-Zener transition is 
\end{subequations}
\begin{equation}
\Delta \mathbf{\hat{I}}_{j}=\boldsymbol{\Gamma }_{j}(T_{f},T_{i})\times 
\mathbf{\hat{I}}_{j}(T_{i}),
\end{equation}%
where the total effect of the electrons on the nuclei is the integrated
effect of the magnetic splitting in Eqs. (\ref{Delta+}), (\ref{Delta-}), and
(\ref{Deltaz}) : 
\begin{equation}
\boldsymbol{\Gamma }_{j}(T_{f},T_{i})=\int_{T_{i}}^{T_{f}}\frac{dt}{\hbar }%
\boldsymbol{\Delta }_{j}(t).  \label{integratedfield}
\end{equation}
%

In order to find explicit expressions for the dependence of the electron
states on the effective field induced by the transverse nuclear spin
polarization $v_{n}^{\pm }$, it is convenient to make a transformation of
the singlet and triplet amplitudes 
\begin{equation}
c_{s}=\tilde{c}_{s}\text{, \  \ }c_{T_{+}}=\tilde{c}_{T_{+}}v^{-}/v_{\bot }.
\label{cstransform}
\end{equation}
%
Then the Hamiltonian becomes real, and 
\begin{equation}
\left( 
\begin{array}{cc}
\epsilon _{s} & v_{\bot } \\ 
v_{\bot } & \epsilon _{T_{+}}%
\end{array}%
\right) \left( 
\begin{array}{c}
\tilde{c}_{S} \\ 
\tilde{c}_{T_{+}}%
\end{array}%
\right) =i\hbar \partial _{t}\left( 
\begin{array}{c}
\tilde{c}_{S} \\ 
\tilde{c}_{T_{+}}%
\end{array}%
\right) ,  \label{Hst+transform}
\end{equation}
%
where $v_{\bot }=\left \vert v^{\pm }\right \vert $. Eq.~(\ref{Hst+transform}%
) depends, in addition to the external magnetic field, on the absolute value
of the combined effect of the nuclear spin induced transverse effective
field and spin-orbit interaction, but does not depend on its direction.

In this basis, we can express the total effect of the $(x,y)$ components of
the effective field of Eq.~(\ref{integratedfield}) in terms of 
\begin{equation}
\Gamma _{j}^{\pm }=\pm iA\rho _{j}\Lambda ^{\pm }v^{\pm }/(2v_{\bot }^{2})\,,
\label{GammaLambda}
\end{equation}%
%
where the dimensionless functions $\Lambda ^{\pm }(T_{i},T_{f})$ are defined
as 
\begin{equation}
\Lambda ^{-}=i2v_{\bot }\int_{T_{i}}^{T_{f}}\frac{dt}{\hbar }\tilde{c}%
_{S}^{\ast }(t)\tilde{c}_{T_{+}}(t),  \label{Lambda}
\end{equation}%
%
and $\Lambda ^{+}=\left( \Lambda ^{-}\right) ^{\ast }$. This expression can
be transformed by using the equation $\tilde{c}_{T_{+}}=v_{\bot
}^{-1}(i\hbar \partial _{t}-\epsilon _{s}(t))\tilde{c}_{s}$ following from
Eq.~(\ref{Hst+transform}), 
\begin{equation}
\Lambda ^{-}=-2\int_{T_{i}}^{T_{f}}dt\tilde{c}_{S}^{\ast }(t)\frac{\partial 
\tilde{c}_{s}(t)}{\partial t}-i2\int_{T_{i}}^{T_{f}}\frac{dt}{\hbar }%
\epsilon _{s}(t)\left \vert c_{s}(t)\right \vert ^{2},  \label{Lambda-}
\end{equation}%
%
so that 
\begin{equation}
\text{Re}\{ \Lambda ^{\pm }\}=P=\left \vert c_{s}(T_{i})\right \vert
^{2}-\left \vert c_{s}(T_{f})\right \vert ^{2},  \label{Probability}
\end{equation}%
%
is the transition probability $P(T_{i},T_{f})$ from the singlet $S$ to the
triplet $T_{+}$ state. There is no such a simple relation between the
imaginary parts of $\Lambda ^{\pm }$ and the transition probability, and
this fact is important for the following discussion of the effect of the
Landau-Zener sweeps on nuclei. However, we observe that when the Hamiltonian
in the Schr\"{o}dinger equation (\ref{Hst+transform}) is stationary, i.e.,
when the gate voltages are fixed and $\epsilon _{S}$ and $\epsilon _{T_{+}}$
are constant in time, and 
the system is in an eigenstate of the Hamiltonian of Eq.~(\ref{Hst+transform}%
), the field $\tilde{c}_{S}^{\ast }\tilde{c}_{T_{+}}$ is real implying a
nonvanishing imaginary contribution to $\Lambda ^{\pm }$. The imaginary part
of $\Lambda ^{\pm }$ thus includes contributions that can be understood in
terms of RKKY-like static nuclear spin-spin interaction mediated by the
electronic state, but this interaction also depends on the spin-orbit
coupling. We will diagonalize the stationary Hamiltonian of Eq.~(\ref%
{Hst+transform}) in Sec. \ref{sec:QS} and relate the imaginary part of $%
\Lambda ^{\pm }$ to the static electronic properties and show how this
influences the dynamical nuclear dynamical properties. The imaginary part of 
$\Lambda ^{\pm }$ is central for the understanding of the dynamical nuclear
polarization and we define%
\begin{equation}
Q=\text{Im}\{ \Lambda ^{+}\}=-\text{Im}\{ \Lambda ^{-}\}.  \label{Qdef}
\end{equation}%
%

We also express the total effect of the field along $z$ as 
\begin{equation}
\Gamma _{j}^{z}=A\zeta _{j}\Lambda ^{z}/(2v_{\bot })-\eta
_{j(nZ)}(T_{f}-T_{i})/\hbar ,  \label{GammazLamdbaz}
\end{equation}
%
where 
\begin{equation}
\Lambda ^{z}=2v_{\bot }\int_{T_{i}}^{T_{f}}\frac{dt}{\hbar }\left \vert 
\tilde{c}_{T_{+}}(t)\right \vert ^{2}.  \label{Gammaz}
\end{equation}

Using Eqs. (\ref{Delta+}), (\ref{Delta-}), and (\ref{Deltaz}), as well as
expressing $\Gamma _{j}^{\pm }$ and $\Gamma _{j}^{z}$ of Eqs. (\ref%
{GammaLambda}) and (\ref{Gammaz}) in terms of $\Lambda ^{\pm }$ and $\Lambda
^{z}$, we arrive at the spin production during a single $S\rightarrow T_+$
transition both in the transverse 
\begin{eqnarray}
\Delta \hat{I}_{j}^{\pm } &=&\frac{A}{2v_{\bot }}\left[ \frac{v^{\pm }}{%
v_{\bot }}\Lambda ^{\pm }\rho _{j}\hat{I}_{j}^{z}\pm i\Lambda ^{z}\zeta _{j}%
\hat{I}_{j}^{\pm }\right]  \notag \\
&&\mp i\frac{\eta_{j(nZ)}}{\hbar }(T_{f}-T_{i})\hat{I}_{j}^{\pm }
\label{deltaIj+-}
\end{eqnarray}
%
and the longitudinal components%
\begin{equation}
\Delta \hat{I}_{j}^{z}=-\frac{A}{2v_{\bot }^{2}}\left[ \Lambda ^{-}v^{-}\rho
_{j}\hat{I}_{j}^{+}+\Lambda ^{+}v^{+}\rho _{j}\hat{I}_{j}^{-}\right] .
\label{deltaIz}
\end{equation}
%
Next, substituting operators $\hat{v}_{n}^{\pm }$ in Eq.~(\ref{vn}) by their
semiclassical values $v_{n}^{\pm }$ and using Eq.~(\ref{vtotal}), we find
the change in the $z$-component of the total nuclear spin, $\Delta
I^{z}=\sum \limits_{j}\Delta I_{j}^{z} $, 
\begin{equation}
\Delta I^{z} =-P+\frac{1}{2v_{\bot }^{2}}\left[ \Lambda
^{-}v^{-}v_{so}^{+}+\Lambda ^{+}v^{+}v_{so}^{-}\right] ,
\label{totalchangeIz}
\end{equation}
or 
\begin{equation}
\Delta I^{z} =-\frac{P}{2v^2_\perp}(v^-v_n^++v^+v_n^-)-i\frac{Q}{2v^2_\perp}%
(v^-v_n^+-v^+v_n^-).  \label{totalchangeIz2}
\end{equation}
Note that the change in the $z$-component of the total nuclear spin is
computed under the constraint that the transverse nuclear fields are $v_n^\pm
$ before the sweep.

Remarkably, $\Delta I^{z}$ of Eq.~(\ref{totalchangeIz}) only depends on the
basic parameters of the Hamiltonian ${H}^{(ST_{+})}$ of Eq.~(\ref{Hst+}) and
the shape of the sweep and does not depend on the detailed topography of
nuclear spins. Therefore, the result is very general and convenient to use.
In this respect, transfer of the longitudinal component of the angular
momentum differs from the transfer of its transverse component that,
according to Eq.~(\ref{deltaIj+-}), depends on the specific spin
configuration.

In the absence of spin-orbit interaction, $v_{so}^{\pm }=0$, the total
change in the electron spin equals the transition probability $P$, as
expected for a (partial)\ transition between the singlet $S$ and triplet $%
T_{+}$ states. Conservation of the $z$ component of the angular momentum
then dictates that the change in the $z$ component of the total nuclear spin
equals $-P$. Spin-orbit coupling breaks the conservation law for the angular
momentum transfer from the electronic to the nuclear spin system since
angular momentum can be transferred to or from the lattice as well. Such
processes manifest themselves in the second term in Eq. (\ref{totalchangeIz}%
). It depends on the relative phase between the spin-orbit and hyperfine
interaction matrix elements. Furthermore, this term depends not only on the
transition probability $P$, 
but also on the imaginary parts of $\Lambda ^{\pm }$. $Q$ 
acquires contributions not only from the part of the sweep near the
anticrossing point but also from its long tails. As a result, the magnitude
of $Q$ can be much larger than $P$ for certain classes of sweeps. This
generic feature suggests that $Q$ can be made large, and the spin-orbit
coupling can strongly influence nuclear dynamics even when it is weaker than
the hyperfine coupling.

Our results confirm the prediction of Ref. \onlinecite{Rudner:prb10b} that
the spin-orbit coupling influences the nuclear spin generation rate
profoundly. The quantity computed in Ref. \onlinecite{Rudner:prb10b} is the
total change of the nuclear spin $\Delta I^{z}$ \textit{averaged} over the
phase of the transverse nuclear field $v_{n}^{\pm }$. This averaging
annihilates the second term of Eq.~(\ref{totalchangeIz2}) while the first
term coincides with Eq. (9) in Ref. \onlinecite{Rudner:prb10b}.\cite%
{RudnerProof} Since $Q$ can be considerably larger than $P$, we expect
enhancement of spin production rate in experiments performed at a fixed
(while generic) values of $v_{n}^{\pm }$.

\section{Linear sweeps - Landau-Zener electron transitions}

\label{sec:LandauZener}


When the changes in the gate voltages are such that the difference in the
energy betwen the singlet $S$ and the triplet $T_{+}$ varies linearly in
time, Eq. (\ref{Hst+transform}) reduces to the standard Landau-Zener
problem. Because the Landau approach based on analytical continuation allows
finding only the transition probabilities,\cite{LL:book65} we employ in the
following the Zener approach\cite{Zener:prs32} allowing finding explicit
expressions for the time dependence of electron wave functions that drives
the coherent nuclear spin dynamics. We consider a transition from the
singlet $S$ state to the triplet $T_{+}$ state, but because of the
symmetries of the Hamiltonian the solution can also be used to find the wave
functions that describe the transition from the triplet $T_{+}$ to the
singlet $S$ state. We derive this relation in Sec.~\ref{sec:reverse}.
Defining $t=0$ as the time when the energies $\epsilon _{s}$ and $\epsilon
_{T_{+}}$ of the singlet $S$ and triplet $T_{+}$ are equal, we introduce 
\begin{equation}
\epsilon _{s}=\beta ^{2}t/2\hbar ,\, \, \, \epsilon _{T_{+}}=-\beta
^{2}t/2\hbar ,  \label{epsilonST}
\end{equation}
%
where $\beta $ is a positive 
number with dimension of energy. This representation implies that the
singlet state has the lowest energy at early (negative) times and the
triplet state has the lowest energy for large final (positive) times. A
natural time-scale is $\hbar /\beta $ so that Eq. (\ref{Hst+transform}) with 
$\tau =t\beta /\hbar $ reads 
\begin{equation}
\left( 
\begin{array}{cc}
\tau /2 & \sqrt{\gamma } \\ 
\sqrt{\gamma } & -\tau /2%
\end{array}%
\right) \left( 
\begin{array}{c}
\tilde{c}_{S} \\ 
\tilde{c}_{T_{+}}%
\end{array}%
\right) =i\partial _{\tau }\left( 
\begin{array}{c}
\tilde{c}_{S} \\ 
\tilde{c}_{T_{+}}%
\end{array}%
\right) ,  \label{LZSchrodinger}
\end{equation}
%
where 
\begin{equation}
\gamma =\left( v_{\bot }/\beta \right) ^{2}  \label{gammaLZ}
\end{equation}
%
is the Landau-Zener parameter. When $\gamma $ is small, the transition
probability from the singlet $S$ to the triplet state $T_{+}$ is small. In
the opposite limit, when $\gamma $ is large, the transition probability is
close to $1$. As above, we denote the initial time from where the 
sweep starts as $T_{i}$ and the final time where it ends as $T_{f}$. In
dimensionless units, we have $\tau _{i}=T_{i}\beta /\hbar $ and $\tau
_{f}=T_{f}\beta /\hbar $.

In order to determine the change in the nuclear spin polarization, we need
to compute not only the transition probability $P$, but also the singlet $S$
and triplet $T_{+}$ amplitudes, $\tilde{c}_{S}$ and $\tilde{c}_{T_{+}}$.
Because the nuclear dynamics is controlled by the electron dynamics via the
effective fields of Eqs. (\ref{Delta+}), (\ref{Delta-}), and (\ref{Deltaz}),
explicit expressions for the amplitudes $(\tilde{c}_{S}(\tau ),\tilde{c}%
_{T_{+}}(\tau ))$ should be found not only near the anticrossing point $\tau
=0$, but along the whole sweep, $\tau _{i}\leq \tau \leq \tau _{f}$.
Therefore, it is necessary to employ Zener's derivation of the Landau-Zener
transition probability \cite{Zener:prs32} and complement it with a detailed
information about the asymptotic behavior of the amplitudes and effective
magnetic fields.

Eliminating $\tilde{c}_{s}$ from Eq.~(\ref{LZSchrodinger}) by substituting 
\begin{equation}
\tilde{c}_{s}=\frac{1}{\sqrt{\gamma }}\left( \frac{\tau }{2}+i\partial
_{\tau }\right) \tilde{c}_{T_{+}},  \label{singletfromtriplet}
\end{equation}%
into its first row, we find 
\begin{equation}
\partial _{\tau }^{2}\tilde{c}_{T_{+}}+\left( \gamma -\frac{i}{2}+\frac{1}{4}%
\tau ^{2}\right) \tilde{c}_{T_{+}}=0.  \label{diffeqct}
\end{equation}
%
Then, by changing the variable $\tau $ to 
\begin{equation}
z=e^{i3\pi /4}\tau ,  \label{ztau}
\end{equation}
%
Eq.~(\ref{diffeqct}) transforms to 
\begin{equation}
\partial _{z}^{2}\tilde{c}_{T_{+}}\left( z\right) +\left( n+\frac{1}{2}-%
\frac{1}{4}z^{2}\right) \tilde{c}_{T_{+}}(z)=0,  \label{diffeqparabolic}
\end{equation}
%
where $n=i\gamma $. This is the Weber equation\cite%
{WhittakerWatson:book45,Erdelyi} whose solutions are the parabolic cylinder
(Weber) functions $D_{n}(z)$, $D_{n}(-z)$, $D_{-1-n}(-iz)$ and $D_{-1-n}(iz)$
of which only two are linearly independent. When expressed as functions of
the real argument $\tau $, they correspond to $D_{i\gamma }(e^{i3\pi /4}\tau
)$, $D_{i\gamma }(-e^{i3\pi /4}\tau )$, $D_{-1-i\gamma }(e^{i\pi /4}\tau )$
and $D_{-1-i\gamma }(-e^{i\pi /4}\tau )$, respectively. In a similar way, we
find the differential equation that the singlet amplitude obeys. Eliminating 
$\tilde{c}_{T_{+}}$ by substituting 
\begin{equation}
\tilde{c}_{T_{+}}=\frac{1}{\sqrt{\gamma }}\left( -\frac{\tau }{2}+i\partial
_{\tau }\right) \tilde{c}_{S},
\end{equation}%
into the second row of Eq.~(\ref{LZSchrodinger}) and taking its complex
conjugate, we find 
\begin{equation}
\partial _{\tau }^{2}\tilde{c}_{S}^{\ast }+\left( \gamma -\frac{i}{2}+\frac{1%
}{4}\tau ^{2}\right) \tilde{c}_{S}^{\ast }=0.  \label{diffeqcs*}
\end{equation}
%
Hence $\tilde{c}_{S}^{\ast }$ satisfies the same differential equation (\ref%
{diffeqct}) as $\tilde{c}_{T_{+}}$; its solutions are the Weber functions
listed above. In Sec.~\ref{sec:AsExp} we discuss the asymptotic behavior of
the singlet $S$ and triplet $T_{+}$ amplitudes that is critical for imposing
the initial conditions and finding long time scale nuclear spin dynamics.

\subsection{Asymptotic Expansions}

\label{sec:AsExp} 

For the following, the asymptotic behavior of the solutions in both limits, $%
\tau \rightarrow \pm \infty $, is required. However, because the solutions
appear in pairs, with opposite signs of $\tau $, it is sufficient to find
their $\tau >0$ asymptotics. We note that the indeces of all above $D$%
-functions are imaginary or complex [$i\gamma $ or $(-1-i\gamma )$] while
the asymptotics of Refs. \onlinecite{WhittakerWatson:book45,Erdelyi} are valid only
for $D_{n}(z)$ functions with integer indeces.\cite{Zenercomment} In what
follows, we employ the asymptotic expressions from Mathematica 8 which are
valid for arbitrary complex indices. For large positive times $\tau
\rightarrow \infty $, they are 
\begin{widetext}
\begin{subequations}
\begin{align}
D_{i\gamma }(e^{i3\pi /4}\tau ) &\approx e^{-3\pi \gamma /4}e^{i\tau
^{2}/4}\tau ^{i\gamma }+e^{i\pi /4}\frac{\sqrt{2\pi }}{\Gamma (-i\gamma )}%
e^{-\pi \gamma /4}e^{-i\tau ^{2}/4}\tau ^{-1-i\gamma }+\mathcal{O}(\tau
^{-2}), \label{asympt1} \\
D_{i\gamma }(-e^{i3\pi /4}\tau ) &\approx  e^{\pi \gamma/4} e^{i\tau ^{2}/4}\tau ^{i\gamma }+%
\mathcal{O}(\tau ^{-2}), \label{asympt2} \\
D_{-1-i\gamma }(e^{i\pi /4}\tau ) &\approx e^{-i\pi /4}e^{\pi \gamma
/4}e^{-i\tau ^{2}/4}\tau ^{-1-i\gamma }+\mathcal{O}(\tau ^{-3}), \label{asympt3} \\
D_{-1-i\gamma }(-e^{i\pi /4}\tau ) &\approx \frac{\sqrt{2\pi }}{\Gamma
(1+i\gamma )} e^{-\pi \gamma/4} e^{i\tau ^{2}/4}\tau ^{i\gamma }+e^{i3\pi /4}e^{-3\pi \gamma
/4}e^{-i\tau ^{2}/4}\tau ^{-1-i\gamma }+\mathcal{O}(\tau ^{-2}). \label{asympt4}
\end{align}
\end{subequations}
\end{widetext}\lbrack asympt1-4] One can see that as $\tau \rightarrow
\infty $ the function $D_{-1-i\gamma }(e^{i\pi \gamma /4}\tau )$ vanishes as 
$\tau ^{-1}$ while the absolute values of the three other D-functions
saturate. We note that all asymptotic expressions for the $D$-functions
include two oscillatory factors. The Fresnel-type factors $\exp (\pm i\tau
^{2}/4)$ originate from the accumulation of the adiabatic Schr\"{o}dinger
phases during a linear sweep, and the factors $\tau ^{\pm i\gamma }$
depending on $\gamma $ reflect the non-adiabaticity.

It follows from Eq.~(\ref{asympt3}) that for a sweep starting from the
singlet $S$ state at large negative initial time $\tau _{i}$, the function $%
D_{-1-i\gamma }(-e^{i\pi \gamma /4}\tau )$ should be chosen as one of the
basis functions for the triplet $T_{+}$ state because it vanishes when $\tau
\rightarrow -\infty $. We choose $D_{i\gamma }(e^{i3\pi \gamma /4}\tau )$ as
the second basis function. Then 
\begin{align}
\tilde{c}_{T_{+}}(\tau )& =a\sqrt{\gamma }e^{-i3\pi /8}D_{-1-i\gamma
}(-e^{i\pi /4}\tau )  \notag \\
& -\frac{b}{\sqrt{\gamma }}e^{-i3\pi /8}D_{i\gamma }\left( e^{i3\pi /4}\tau
\right) ,  \label{cT+general}
\end{align}
%
where $a$ and $b$ are 
coefficients that depend on the initial time $\tau _{i}$. The overall phase
factor as well as the factors $\sqrt{\gamma }$ and $-1/\sqrt{\gamma }$ have
been chosen as a matter of convenience in the following transformation. One
can check that $b\propto \tau _{i}^{-2}$ for $|\tau _{i}|\gg 1$.

Eq.\ (\ref{diffeqcs*}) implies that $\tilde{c}_{S}^{\ast }$, the complex
conjugate of the singlet $S$ amplitude, can be expressed in terms of the
same Weber functions as the triplet amplitude ${\tilde{c}}_{T_{+}}$. An
explicit connection between them can be found by employing Eq. (\ref%
{singletfromtriplet}), and the expression for the singlet component $\tilde{c%
}_{s}$ can be further simplified by using the standard recurrence relations
for $D$-functions.\cite{WhittakerWatson:book45,Erdelyi} As applied to the $D$%
-functions of Eq. (\ref{cT+general}), they read 
\begin{equation}
\left( \frac{\tau }{2}+i\partial _{\tau }\right) D_{i\gamma }(e^{i3\pi
/4}\tau )=-\gamma e^{i3\pi /4}D_{-1+i\gamma }(e^{i3\pi /4}\tau )
\label{Dcomplex1}
\end{equation}%
%
and 
\begin{equation}
\left( \frac{\tau }{2}+i\partial _{\tau }\right) D_{-1-i\gamma }(-e^{i\pi
/4}\tau )=e^{i3\pi /4}D_{-i\gamma }(-e^{i\pi /4}\tau )  \label{Dcomplex2}
\end{equation}%
%
The $D$-functions of the right hand side of Eqs. (\ref{Dcomplex1}) and (\ref%
{Dcomplex2}) differ from the D-functions of Eq. (\ref{cT+general}), but are
related to them by complex conjugation 
\begin{eqnarray}
D_{-1+i\gamma }(e^{i3\pi /4}\tau ) &=&\left[ D_{-1-i\gamma }(-e^{i\pi
/4}\tau )\right] ^{\ast }, \\
D_{-i\gamma }(-e^{i\pi /4}\tau ) &=&\left[ D_{i\gamma }(e^{i3\pi /4}\tau )%
\right] ^{\ast }.
\end{eqnarray}

Therefore, the general solution for the singlet amplitudes is 
\begin{eqnarray}
\tilde{c}_{S}\left( \tau \right) &=&a\left[ e^{-i3\pi /8}D_{i\gamma
}(e^{i3\pi /4}\tau )\right] ^{\ast }  \notag \\
&&+b\left[ e^{-i3\pi /8}D_{-1-i\gamma }(-e^{i\pi /4}\tau )\right] ^{\ast }\,.
\label{cSgeneral}
\end{eqnarray}%
As a consequence, the function $\Lambda ^{-}$ of Eq. (\ref{Lambda})
depending on the product $\tilde{c}_{S}^{\ast }(t)\tilde{c}_{T_{+}}(t)$ and
describing the response of nuclear spins to a Landau-Zener pulse can be
expressed in terms of two functions $D_{-1-i\gamma }(-e^{i\pi \gamma /4}\tau
)$\ and $D_{i\gamma }(e^{i3\pi \gamma /4}\tau )$. In Sec.~\ref%
{sec:infinitelimits}, we consider the Landau-Zener scenario when the initial
electron state is prepared at $\tau _{i}\rightarrow -\infty $ and the sweep
runs to $\tau_f\rightarrow \infty$, as well as the asymptotic behavior of
effective fields ${\tilde c}^*_S{\tilde c}_{T_+}$ at large but finite times $%
|\tau|\gg1$.

\subsection{Infinite Limits and Asymptotics}

\label{sec:infinitelimits}


When the system is in the singlet state at early times, $\left \vert \tilde{c%
}_{S}(\tau \rightarrow -\infty )\right \vert =1$ and $\tilde{c}_{T_{+}}(\tau
\rightarrow -\infty )=0$, then $b=0$ and $\left \vert a\right \vert
^{2}e^{\pi \gamma /2}=1$, as follow from Eq.~(\ref{asympt2}), and 
\begin{subequations}
\label{cSandcT+}
\begin{align}
\tilde{c}_{S}(\tau )& =e^{i\varphi }e^{-\pi \gamma /4}\left[ e^{-i3\pi
/8}D_{i\gamma }(e^{i3\pi /4}\tau )\right] ^{\ast },  \label{cS} \\
\tilde{c}_{T_{+}}(\tau )& =e^{i\varphi }e^{-\pi \gamma /4}\sqrt{\gamma }%
\left[ e^{-i3\pi /8}D_{-1-i\gamma }(-e^{i\pi /4}\tau )\right] ,  \label{cT+}
\end{align}
%
where $\varphi $ is an arbitrary phase. For a finite but large initial time $%
-\tau _{i}$ ($\tau _{i}>0$), this description remains satisfactory with the
accuracy to the terms of the order $\tau _{i}^{-2}$ in the singlet amplitude
of Eq. (\ref{cS}) and of the order $\tau _{i}^{-1}$ in the triplet amplitude
of Eq. (\ref{cT+}).

For completeness, let us also consider the situation when the system is in
the triplet state $T_{+}$ at early times $\tau \rightarrow -\infty $. Then
it follows from Eqs.\ (\ref{asympt2}) and (\ref{asympt3}) that $a=0$ and $%
e^{\pi \gamma /2}|b|^{2}/\gamma =1$, so that 
\end{subequations}
\begin{subequations}
\label{cSandcT+tripinit}
\begin{align}
\tilde{c}_{S}(\tau )& =e^{i\varphi ^{\prime }}e^{-\pi \gamma /4}\sqrt{\gamma 
}\left[ e^{-i3\pi /8}D_{-1-i\gamma }(-e^{i\pi /4}\tau )\right] ^{\ast },
\label{cStripinit} \\
\tilde{c}_{T_{+}}(\tau )& =-e^{i\varphi ^{\prime }}e^{-\pi \gamma /4}\left[
e^{-i3\pi /8}D_{i\gamma }(e^{i3\pi /4}\tau )\right] ,  \label{cT+tripinit}
\end{align}
%
where $\phi ^{\prime }$ is an arbitrary phase.

We can now find the transition probability for the $S\rightarrow T_{+}$
transition of Eq.~(\ref{Probability}). It is

\end{subequations}
\begin{equation}
P_{LZ}=\left \vert \tilde{c}_{S}(\tau \rightarrow -\infty )\right \vert
^{2}-\left \vert \tilde{c}_{S}(\tau \rightarrow \infty )\right \vert
^{2}=1-e^{-2\pi \gamma }.  \label{PLandauZener}
\end{equation}%
%
which is the celebrated Landau-Zener result. The transverse components of
the effective field acting on the nuclear spins are controlled by the product%
\begin{eqnarray}
\tilde{c}_{S}^{\ast }\tilde{c}_{T_{+}} &=&\sqrt{\gamma }e^{-\pi \gamma
/2}e^{-i3\pi /4}\times   \label{effectivefieldinfinite} \\
&&D_{i\gamma }(e^{i3\pi /4}\tau )D_{-1-i\gamma }(-e^{i\pi /4}\tau ).  \notag
\end{eqnarray}%
Its asymptotic behavoir following from Eqs. (\ref{asympt2}) and (\ref%
{asympt4}) is%
\begin{equation}
\tilde{c}_{S}^{\ast }\tilde{c}_{T_{+}}\approx \frac{\sqrt{\gamma }}{\tau }+%
\mathcal{O}(\tau ^{-2})  \label{sconjugatetnegativeasympt}
\end{equation}%
%
for the early times $\tau \rightarrow -\infty $ and 
\begin{widetext}
\begin{equation}
\tilde{c}_{S}^{\ast }\tilde{c}_{T_{+}}\approx -\frac{\sqrt{\gamma }}{\tau }%
\left[ 1-2e^{-2\pi \gamma }\right] +\sqrt{\gamma}  e^{-i3\pi /4}\frac{\sqrt{2 \pi}}{\Gamma (1+i\gamma )}e^{- 3\pi \gamma /2}e^{i\tau ^{2}/2}\tau ^{2i\gamma } +\mathcal{O}%
(\tau ^{-2})  \label{sconjugatetpositiveasympt}
\end{equation}%
\end{widetext}for the late times $\tau \rightarrow \infty $. The absolute
value of the second term of Eq. (\ref{sconjugatetpositiveasympt}) is $%
e^{-\pi \gamma }\sqrt{1-e^{-2\pi \gamma }}$ as can be checked by using the
identity $\mid \Gamma (1+i\gamma )\mid ^{2}=\pi \gamma /\sinh (\pi \gamma )$%
. This result is easy to understand since it equals $|\tilde{c}_{S}||\tilde{c%
}_{T_{+}}|$ in the asymptotic regime $\tau \rightarrow \infty $, where $|%
\tilde{c}_{T_{+}}|^{2}=1-e^{-2\pi \gamma }$ and $|\tilde{c}%
_{S}|^{2}=e^{-2\pi \gamma }$. The second term of Eq.~(\ref%
{sconjugatetpositiveasympt}) exhibits very fast Fresnel-like oscillations $%
e^{i\tau ^{2}/2}$ when $\tau \rightarrow \infty $ and does not contribute
significantly to the integral $\Lambda ^{-}$ of Eq. (\ref{Lambda})
describing the total effective field applied to the nuclei as a result of
the sweep. This factor originates from the accumulation of the phase $\exp
\left \{ \int \left[ \epsilon _{S}(t)-\epsilon _{T_{+}}(t)\right] dt/\hbar
\right \} $ along the sweep.

The origin of the coefficients in the $1/\tau $ terms in Eqs. (\ref%
{sconjugatetnegativeasympt}) and (\ref{sconjugatetpositiveasympt}) can also
be made quite transparent. By using the time-dependent Schr\"{o}dinger
equation (\ref{LZSchrodinger}), we find 
\begin{equation}
\left( i\partial _{\tau }+\tau \right) \left( \tilde{c}_{S}^{\ast }\tilde{c}%
_{T_{+}}\right) =\sqrt{\gamma }\left[ \left \vert \tilde{c}_{S}\left( \tau
\right) \right \vert ^{2}-\left \vert \tilde{c}_{T_{+}}\left( \tau \right)
\right \vert ^{2}\right] \cdot
\end{equation}%
Knowing that for early times, $\tau \rightarrow -\infty $, the amplitudes
approach $\left \vert \tilde{c}_{S}\right \vert ^{2}=1$ and $\left \vert 
\tilde{c}_{T_{+}}\right \vert ^{2}=0$, we recover Eq. (\ref%
{sconjugatetnegativeasympt}). \ For late times, $\left \vert \tilde{c}%
_{S}\right \vert ^{2}-\left \vert \tilde{c}_{T_{+}}\right \vert
^{2}\rightarrow -1+2\exp (-2\pi \gamma )$, which explains the $1/\tau $ term
in Eq. (\ref{sconjugatetpositiveasympt}). Furthermore, we note that in the
leading order the operator $(i\partial _{\tau }+\tau )$ annihilates the
second term of Eq. (\ref{sconjugatetpositiveasympt}).

The integrals of Eqs. (\ref{sconjugatetnegativeasympt}) and (\ref%
{sconjugatetpositiveasympt}) diverge logarithmically when the integration
limits approach $\pm \infty $. This means that while $P_{LZ}$ of Eq. (\ref%
{PLandauZener}) and the total spin transfer $\Delta I^{z}$ of Eq. (\ref%
{totalchangeIz}) (for $v_{so}{}^{\pm }=0$) are controlled by the vicinity of
the anticrossing point, the effective fields $\boldsymbol{\Delta }_{j}$ and
shake up processes in the nuclear subsystem produced by them are controlled
by the global shape of the pulse. The same is true for $\Delta I^{z}$ when $%
v_{so}{}^{\pm }\neq 0$. We note that while the presence of logarithmic terms
is a general property of linear sweeps, they contribute to $\Delta I^{z}$
only in the presence of spin-orbit coupling.

\subsection{Reverse sweep from the triplet $T_{+}$ to the singlet $S$.}

\label{sec:reverse} 

Let us relate the reverse sweep, starting in a triplet state $T_{+}$ and
sweeping to a singlet state $S$, to the $S\rightarrow T_{+}$ sweep
elaborated above. Since now the rates of the change of the singlet $S$ and
triplet $T_{+}$ energies have the signs opposite to the signs in Eq.~(\ref%
{epsilonST}), the dynamical equations for the amplitudes $(\tilde{c}_{S},{%
\tilde{c}}_{T_{+}})$ differ from Eq.~(\ref{Hst+transform}) by the
interchange $\tilde{c}_{S}\leftrightarrow \tilde{c}_{T+}$. Furthermore, for
a $T_{+}\rightarrow S$ transition, the system was initially in the triplet $%
T_{+}$ state, hence, the singlet $S$ amplitude vanishes at the early time.
Therefore, the initial conditions are also $\tilde{c}_{S}\leftrightarrow 
\tilde{c}_{T+}$ interchanged as compared to the $S\rightarrow T_{+}$ sweep.
This implies that their product transforms as $\tilde{c}_{S}^{\ast }\tilde{c}%
_{T_{+}}\rightarrow \left( \tilde{c}_{S}^{\ast }\tilde{c}_{T_{+}}\right)
^{\ast }$, and $\Lambda ^{\pm }\rightarrow -\left( \Lambda ^{\pm }\right)
^{\ast }$according to Eq. (\ref{Lambda}\}. In other words the transition
probability $P=\text{Re}\{ \Lambda ^{\pm }\}$ changes sign, but the
imaginary parts $Q=\text{Im}\left \{ \Lambda ^{+}\right \} $ remain
unchanged. The change of the sign of $\text{Re}\{ \Lambda ^{\pm }\}$ is
obvious because of the $S\leftrightarrow T_{+}$ interchange, so that the
longitudinal component of the angular momentum transfer changes sign.
However, the effective field $\text{Im}\left \{ {\Delta }^{\pm }\right \} $
does not change, and this indicates that the imaginary components of $%
\Lambda ^{\pm }$ should add during a $S\rightarrow T_{+}\rightarrow S$ cycle.

In conclusion of this section, for linear sweeps 
the dimensionless function $\Lambda ^{-}(T_{i},T_{f})$ that reflects the
effect of a single Landau-Zener sweep on nuclei diverges logarithmically
when $T_{i}\rightarrow -\infty $ and $T_{f}\rightarrow \infty $. In Sec.~\ref%
{sec:STS}, we discuss in more detail the dependence of $\Lambda ^{\pm
}(T_{i},T_{f})$ on the limits $(T_{i},T_{f})$ and the Landau-Zener parameter 
$\gamma $.

\subsection{Adiabatic Regime}

\label{sec:QS}


Some more insight on the long-$\tau $ tails of the products ${\tilde{c}}%
_{S}^{\ast }{\tilde{c}}_{T_{+}}$ comes from the stationary solution of Eq. (%
\ref{Hst+transform}).\cite{Taylor:prb07} For a large detuning $\delta
=\epsilon _{T_{+}}-\epsilon _{S}$ from the $S-T_{+}$ anticrossing, when $%
\left \vert \tau \right \vert \gg 1$, the stationary solution of Eq. (\ref%
{Hst+transform}) provides an adiabatic approximation to the singlet and
triplet amplitudes. Note that we still assume the duration of the sweep is
short as compared to the nuclear Larmor precession time.

Then the eigenenergies of the electronic states of the Hamiltonian of Eq.~(%
\ref{Hst+transform}) are 
\begin{equation}
\epsilon _{\pm }=\frac{1}{2}\left( \epsilon _{s}+\epsilon _{T_{+}}\right)
\pm \sqrt{v_{\perp }^{2}+\left( \delta /2\right) ^{2}},
\end{equation}%
and at the lower branch of the energy spectrum the product of the amplitudes
equals 
\begin{equation}
{\tilde{c}}_{S}^{\ast }{\tilde{c}}_{T_{+}}=-\frac{v_{\perp }/2}{\sqrt{%
v_{\perp }^{2}+(\delta /2)^{2}}}.  \label{QS.1}
\end{equation}
%
Here the oscillatory $\tau $-dependent phase factors cancel betause ${\tilde{%
c}}_{S}$ and ${\tilde{c}}_{T_{+}}$ belong to the same eigenvalue. It
immediately allows calculating the transverse components $\Delta
_{j}^{+}=A\varrho _{j}{\tilde{c}}_{S}{\tilde{c}}_{T_{+}}^{\ast }v^{\pm
}/v_{\perp }$ and $\Delta _{j}^{-}=A\varrho _{j}{\tilde{c}}_{S}^{\ast }{%
\tilde{c}}_{T_{+}}v^{-}/v_{\perp }$ of $\boldsymbol{\Delta }_{j}$ and the
effective fields from Eqs.~(\ref{Delta+}) and (\ref{Delta-}). The transverse
components $\Delta _{j}^{\pm }$ vanish as $v_{\perp }/\delta $ when $%
\left
\vert \delta \right \vert /v_{\perp }$ $\rightarrow \infty $.
Similarly, the longitudinal component found from Eq.~(\ref{Deltaz}) equals 
\begin{equation}
\Delta _{j}^{z}=-\frac{A\zeta _{j}}{2}\left[ 1-\frac{\delta /2}{\sqrt{%
v_{\perp }^{2}+\left( \delta /2\right) ^{2}}}\right] -\eta _{j(nZ)}.
\end{equation}%
Far from the intersection, when $\delta /v_{\perp }\rightarrow -\infty $ and
the eigenstate is almost a pure triplet, $\Delta _{j}^{z}\rightarrow -A\zeta
_{j}-\eta _{j(nZ)}$. In the opposite limit, when $\delta /v_{\perp
}\rightarrow \infty $ and the eigenstate is almost a pure singlet, $\Delta
_{j}^{z}\rightarrow -\eta _{j(nZ)}$. The point $\delta =0$ has been
identified as ``spin funnel" in Ref. \onlinecite{Petta2008}.

In the adiabatic limit, the fields $\mbox{\boldmath$\Delta$}_j$ acquire the
usual meaning of RKKY fields with a nuclear dynamic time scale of $%
t\sim \hbar/\Delta_j$. Near the level anticrossing point $\delta=0$, $%
\Delta_j\sim An_0/N$ where $n_0$ is the concentration of nuclei and $N$ is
the number of nuclei in the dot. With $An_0\approx10^{-4}$ eV and $%
N\approx10^6$, $t\approx10 \mu$s.

For a slow linear sweep between $\tau _{i}=-\tau _{f}$ and $\tau _{f}$, with 
$\delta \rightarrow \beta \tau $, one finds from Eqs.~(\ref{Lambda}) and (%
\ref{QS.1}) the quantity $\Lambda _{(a)}^{\pm }$ which, according to Eq. (%
\ref{Qdef}), result in 
\begin{equation}
Q_{(a)}=4\gamma \ln \left( \frac{\sqrt{\tau _{f}^{2}+4\gamma }+\tau _{f}^{2}%
}{2\sqrt{\gamma }}\right) ,  \label{QS.2}
\end{equation}
%
and from Eq. (\ref{Probability}) we find $P_{(a)}=0$. The results for $%
P_{(a)}$ and $Q_{(a)}$ hold with logarithmic accuracy; the subscript $(a)$
indicates that they were derived in the adiabatic approximation. In the same
way, one can check that ${\tilde{c}}_{S}^{\ast }{\tilde{c}}_{T_{+}}$ of Eq.~(%
\ref{QS.1}) is in agreement with the $1/\tau $ terms of Eqs.~(\ref%
{sconjugatetnegativeasympt}) and (\ref{sconjugatetpositiveasympt}).

Applying Eq.~(\ref{QS.1}) to a nonlinear dependence $\delta =\delta (\tau )$%
, one easily concludes that $\Lambda ^{\pm }$ converges if $\delta (\tau )$
is superlinear and diverges by some power law if it is sublinear.

Equation (\ref{QS.1}) implies important consequences for the nuclear spin
dynamics under the condition of time-independent detuning. Indeed, it
follows from Eqs.~(\ref{vtotal}), (\ref{NucDyn}) -- (\ref{Delta-}), and (\ref%
{QS.1}) that the rate of change of the total nuclear spin is 
\begin{equation}
\hbar \frac{\partial I^{z}}{\partial t}=-\frac{i}{2}\frac{%
v_{so}^{+}v_{n}^{-}-v_{so}^{-}v_{n}^{+}}{\sqrt{v_{\perp }^{2}+(\delta /2)^{2}%
}}.  \label{QS.3}
\end{equation}%
%
Therefore, time-independent detuning results in producing a magnetization $%
I^{z}$ that increases linearly in time as long as the parameters of the
electronic Hamiltonian remain unchanged. This generation of spin
magnetization by time-independent electrical bias is possible because the
time-inversion symmetry is violated by a strong external field $\mathbf{B}$
producing Zeeman splitting of the electron triplet state, and the
simultaneous presence of hyperfine and spin-orbit interactions. The
magnitude of the effect reaches its maximum at $\delta =0$, when the system
is brought to the center of the $ST_{+}$ anticrossing. The time scales of
the parameter change can be estimated similarly to Sec. \ref{sec:VIID}.
Under the usual conditions, the shortest of them corresponds to the
precession of $v_{n}^{\pm }$ in the external field. These conclusions seem
to agree with the observations of Ref.~\onlinecite{Foletti2008}.

\section{$S\rightarrow T_+$ sweeps and round cycles}

\label{sec:STS}


Complex functions $\Lambda ^{\pm }(T_{i},T_{f})$ of Eq. (\ref{Lambda})
describe the effect of a sweep on the nuclear spins. As seen from Eqs.~(\ref%
{Probability}) and (\ref{totalchangeIz}), the probability of the electron $%
S\rightarrow T_{+}$ transition $P$ is completely controlled by the real part
of $\Lambda ^{\pm }$, $P=\mathrm{Re}\{ \Lambda ^{\pm }\}$, while the angular
momentum transfered to the nuclear system $\Delta I^{z}$ depends both on the
real and imaginary parts of $\Lambda ^{\pm }$. Imaginary parts of $\Lambda
^{\pm }$ are always present but manifest themselves in the nuclear spin
accumulation only when there are two competing mechanisms of the electron
spin transfer, hyperfine and spin-orbit.

In this section, we first present data on the dependence of $\Lambda ^{\pm }$
on the integration limits and the Landau-Zener parameter $\gamma $ obtained
by numerical integration of Eq.~(\ref{LZSchrodinger}), and then develop an
analytical approach for describing the oscillatory dependence of the
transition probability $P$ on the cycle length.

\subsection{Linear sweeps}

\label{sec:linear}


We begin with linear $S\rightarrow T_{+}$ sweeps of Sec. \ref%
{sec:LandauZener}. For such sweeps, we denote the initial time $-\tau _{i}$ (%
$\tau _{i}>0$) and the final time $\tau _{f}$ ($\tau _{f}>0$) so that the
duration of the sweep is $\tau _{i}+\tau _{f}$. To reduce the number of
parameters, we assume $\tau _{i}=\tau _{f}$. Transition probabilities $%
P(\tau _{f})$ are plotted in Fig.~2(a) as a function of the sweep half-time $%
\tau _{f}$ for two values of $\gamma $. While for large $\tau _{f}$ both
curves saturate to the Landau-Zener probabilities $P_{LZ}$ of Eq. (\ref%
{PLandauZener}), oscillations of $P(\tau _{f})$ are very pronounced. They
decay at a rather long time scale, and their shape cannot be described by a
single characteristic time. We attribute the oscillations to the
interference pattern between two spectrum branches and estimate their period 
$\tau _{osc}$ from the Schr\"{o}dinger exponent $\exp (-iv_{\perp }t/\hbar )$
in the anticrossing point, what results in $\tau _{osc}\approx \gamma
^{-1/2} $. The rate of their decay is controlled by the passage time $\hbar
v_{\perp }/\beta ^{2}$ across the avoided crossing that results in a decay
time $\tau _{dec}\approx \gamma ^{1/2}$. Finally, we arrive at a rough
estimate of the transient regime $\tau _{tr}\sim \max \{{\gamma^{1/2},\gamma
^{-1/2}}\}$. Actually, this only is a lower bound on $\tau _{tr}$. The
saturation takes a longer time and the difference in the shapes of the $%
\gamma =1$ and $\gamma =0.1$ curves deserves more comments. The $\gamma =0.1$
curve strongly resembles plots of Fresnel integrals, and we attribute the
oscillatons to the $e^{i\tau ^{2}/2}$ factors in the asymptotics of Eq. (\ref%
{sconjugatetpositiveasympt}). With increasing $\gamma $, the patterns of
oscillations are getting less regular due to the second oscillatory factor $%
\tau ^{2i\gamma }$ in the asymptotics of $\tilde{c}_{s}^{\ast }\tilde{c}%
_{T_{+}}$. The switching of regimes happens at $2\pi \gamma \approx 1$ as is
seen from the expression $e^{-2\pi \gamma }$ for the Landau-Zener transition
probability. 
\begin{figure}[tbph]
\centering
\subfigure[]{
\includegraphics[width=0.9\columnwidth]{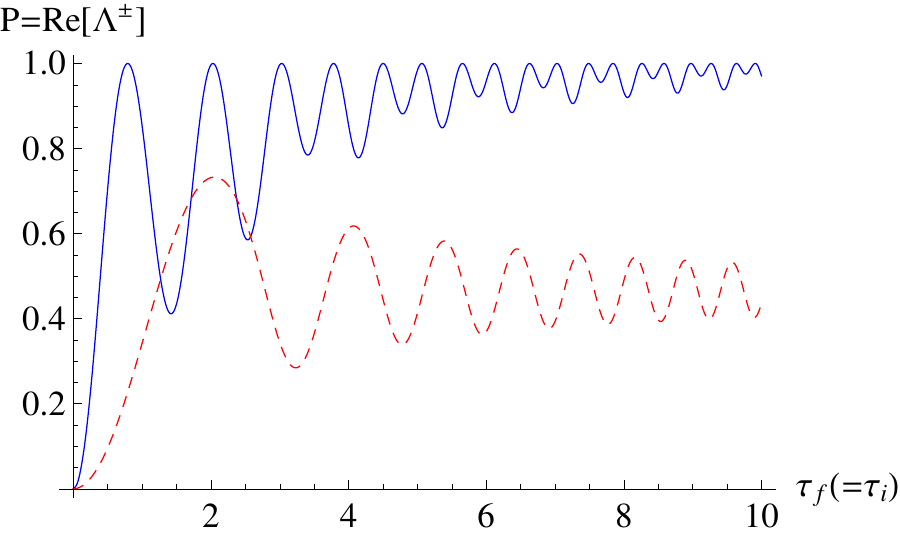}
\label{fig:problinear}
} 
\subfigure[]{
\includegraphics[width=0.9\columnwidth]{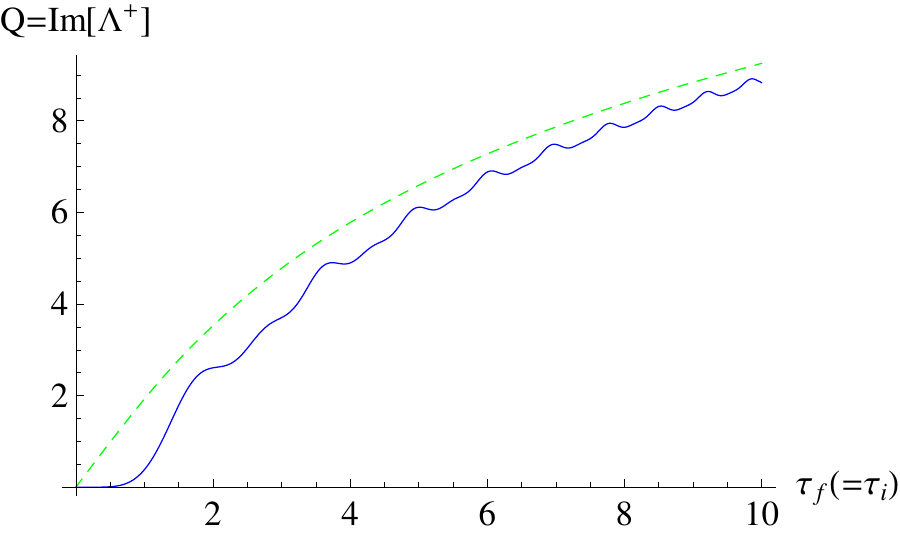}
\label{fig:ImLambdalinearquasistaticgamma1}
} 
\subfigure[]{
\includegraphics[width=0.9\columnwidth]{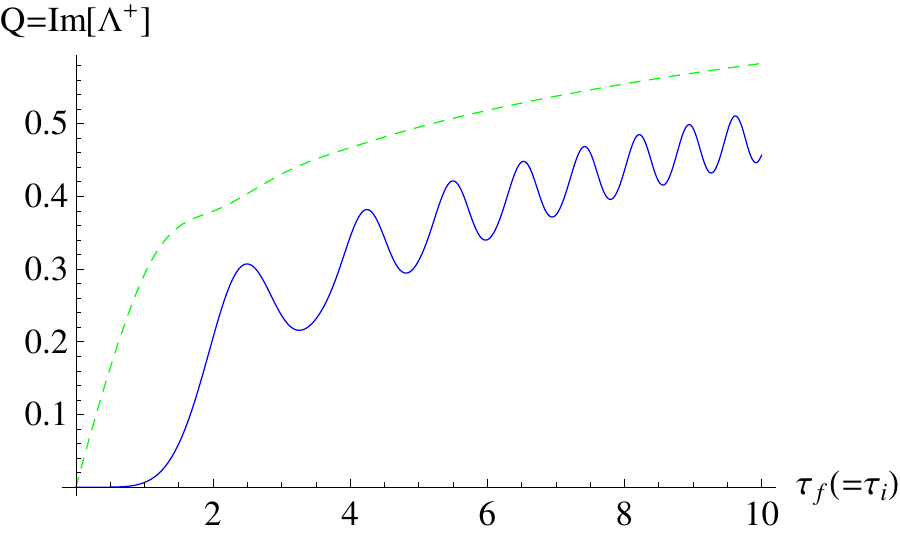}
\label{fig:ImLambdalinearquasistaticgamma01}
}
\caption{(a) Transition probability $P=\text{Re} \{ \Lambda ^{\pm }(\protect%
\tau _{f})\}$ for a linear sweep starting in the $S$ state at the initial
time $-\protect \tau _{i}$ and ending at the final time $\protect \tau _{f}=%
\protect \tau _{i}$ plotted as a function of the half-sweep time $\protect%
\tau _{f}$ for two values of the Landau-Zener parameter $\protect \gamma $.
Full (blue) line $\protect \gamma =1$, dashed (red) line $\protect \gamma =0.1$%
. The anticrossing point is passed in the middle of the sweep at time $%
\protect \tau =0$. The full (blue) lines in (b) and (c) are $Q=\text{Im}%
\{ \Lambda ^{+}(\protect \tau _{f})\}$ for $\protect \gamma =1$ and $\protect%
\gamma =0.1$, respectively. In (b) and (c), the dashed (green) lines are the
adiabatic solutions of Eq.~(\protect \ref{QS.2}).}
\label{fig:Lambdalinear}
\end{figure}

In agreement with the asymptotics found in Sec. \ref{sec:infinitelimits},
the imaginary parts of $\Lambda ^{\pm }$ displayed in Fig.~2(b,c) exhibit a
behavior quite different from the behavior of their real parts $P$. They
increase nearly logarithmically with $\tau_f$, with weak oscillations
superimposed on this monotonic growth. Their magnitudes increase with $\gamma
$, and for $\gamma \approx1$ and $\tau_f\approx10$ they are by one order of
magnitude larger than $P$. Therefore, even with a moderate spin-orbit
coupling, the imaginary parts of $\Lambda ^{\pm }$ are expected to
contribute essentially to the spin transfer $\Delta I^z$ of Eq.~(\ref%
{totalchangeIz}). This contribution should not only change the magnitude of $%
\Delta I^z$ but also smoothen its $\tau_f$-dependence.

In Fig. \ref{fig:ImLambdalinearquasistaticgamma1}, we also plot $Q=\text{Im}%
\{ \Lambda ^{+}\}$ for $\gamma =1$ by using the approximate adiabatic
expression of Eq. (\ref{QS.2}) to compare it to the exact numerical result.
Apart from some details of the behavior for early and late times, which are
expected, we see that the dominant contribution to $Q$ can be explained in
terms of the adiabatic field of Eq. (\ref{QS.2}). Fig. \ref%
{fig:ImLambdalinearquasistaticgamma01} provides a similar comparison, but
for a faster sweep with $\gamma =0.1$. Even in this situation, the adiabatic
approximation is a reasonable starting point for describing the basic shape
of $Q$ of Eq. \ref{Qdef}.

The above analysis of linear sweeps, together with the arguments of Sec.~\ref%
{sec:QS}, allow to make some conclusions about the generic (non-linear) $S$-$%
T_{+}$ sweeps as well. Imagine the sweeps with the rate unchanged near the
anticrossing but increasing away from it. As long as the speed-up happens at
times $\tau >\tau _{tr}$ (this inequality should be fulfilled strong
enough), the probability $P=\text{Re}\{ \Lambda ^{\pm }\}$ changes only
modestly, while the long time tails of the products ${\tilde{c}}_{S}^{\ast
}(\tau ){\tilde{c}}_{T_{+}}(\tau )$ contributing to $Q=\text{Im}\{ \Lambda
^{+}\}$ are cut-off. Thus, increasing the sweep rate away from the
anticrossing reduces $Q$ and might have a profound effect on $\Delta I^{z}$.
However, its specific magnitude depends on the values of a number of
parameters such as $v_{n}^{\pm },v_{so}^{\pm },\tau _{tr}$, and the speed-up
time.

\subsection{Cyclic linear sweeps}

\label{sec:cyclic}


Round sweeps are of the most practical interest for experiment, and their
detailed shapes are nontrivial because of the oscillating tails of $\mathrm{%
Re}\{ \Lambda ^{\pm }\}$ of Fig.~2(a). Therefore, we provide below the data
on $\Lambda ^{\pm }$ for two different round sweeps starting in the singlet
states $S$ at $\tau _{i}<0$.

\begin{figure}[tbph]
\centering
\subfigure{
\includegraphics[width=0.9\columnwidth]{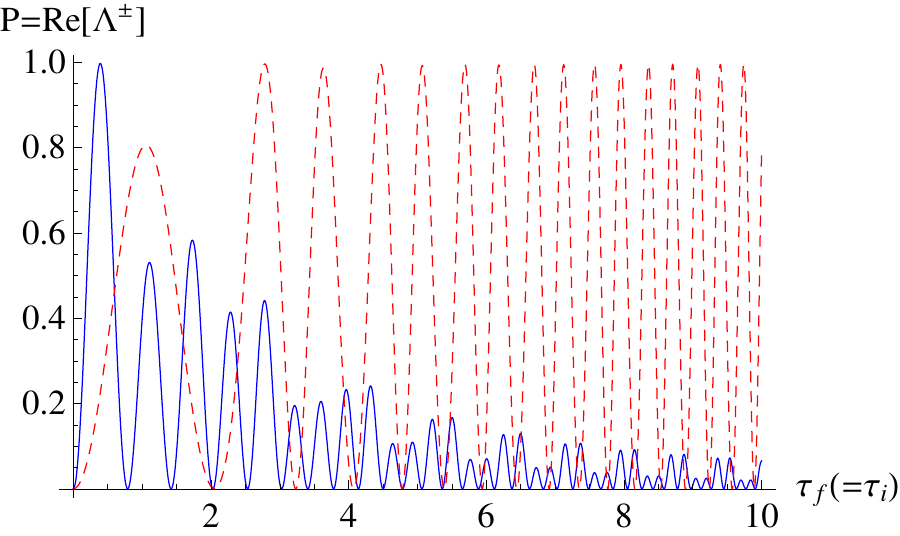}
} 
\subfigure{
\includegraphics[width=0.9\columnwidth]{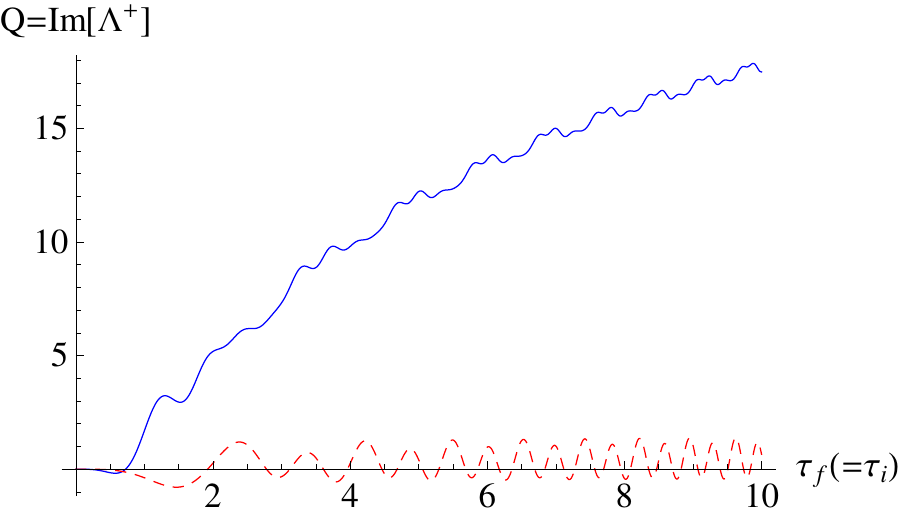}
}
\caption{(a) Transition probability of a $S\rightarrow T_{+}$ transition $P=%
\text{Re}\{ \Lambda ^{\pm }\}$ and (b) the imaginary part $Q=\text{Im}%
\{ \Lambda ^{+}(\protect \tau _{f})\}$ for a round sweep plotted versus $%
\protect \tau_{f}$ (one fourth of the sweep time). The first part of the
sweep is the same as the sweep of Fig.~\protect \ref{fig:Lambdalinear}, and
the second part sweeps in the opposite direction with the same speed
immediately after reaching the turning point. Full (blue) lines $\protect%
\gamma =1$, dashed (red) lines $\protect \gamma =0.1$.}
\label{fig:Lambdalinearbackandforth}
\end{figure}

Fig.~\ref{fig:Lambdalinearbackandforth} presents data for a round sweep of
the total duration of $4\tau _{f}$ that includes the sweep of Fig.~2 from $%
\tau _{i}=-\tau _{f}$ to $\tau _{f}$ and the backward sweep that begins
immediately after the end of the forward sweep. According to Eq.~(\ref%
{Probability}), $P=\mathrm{Re}\{ \Lambda ^{\pm }\}$ displays the probability
of $S\rightarrow T_{+}$ transition. Remarkably, Fig.~3(a) shows that for $%
\gamma =1$ the decay of $P$ is rather long and includes deep and irregular
oscillations. For $\gamma =0.1$, $P(\tau _{f})$ shows a wide maximum at $%
\tau _{f}\approx 2$, and the following oscillations without any visible
decay up to $\tau _{f}=10$. In this case, a double dot in the linear sweep
regime resembles a resonator of a length decreasing as $\tau _{f}^{-1}$. We
expect that first peaks can be resolved experimentally, e.g., in beam
splitter experiments\cite{Petta2010} while higher peaks should merge into a
background with $P\approx 0.5$. Using first sharp peaks for ultrafast spin
operation is highly tempting. 

As distinct from $P=\mathrm{Re}\{ \Lambda ^{\pm }\}$, $Q=\mathrm{Im}\{
\Lambda ^{+}\}$ of Fig.~3(b) is a nearly monotonic function of $\tau _{f}$
for $\gamma =1$ (with irregular oscillations superimposed), and is about 10
for $\tau _{f}=10$. Therefore, it can heavily contribute to $\Delta I^{z}$.
However, $\mathrm{Im}\{ \Lambda ^{\pm }\}$ is small and strongly oscillates
at $\gamma =0.1$.

\begin{figure}[tbph]
\centering
\subfigure{
\includegraphics[width=0.9\columnwidth]{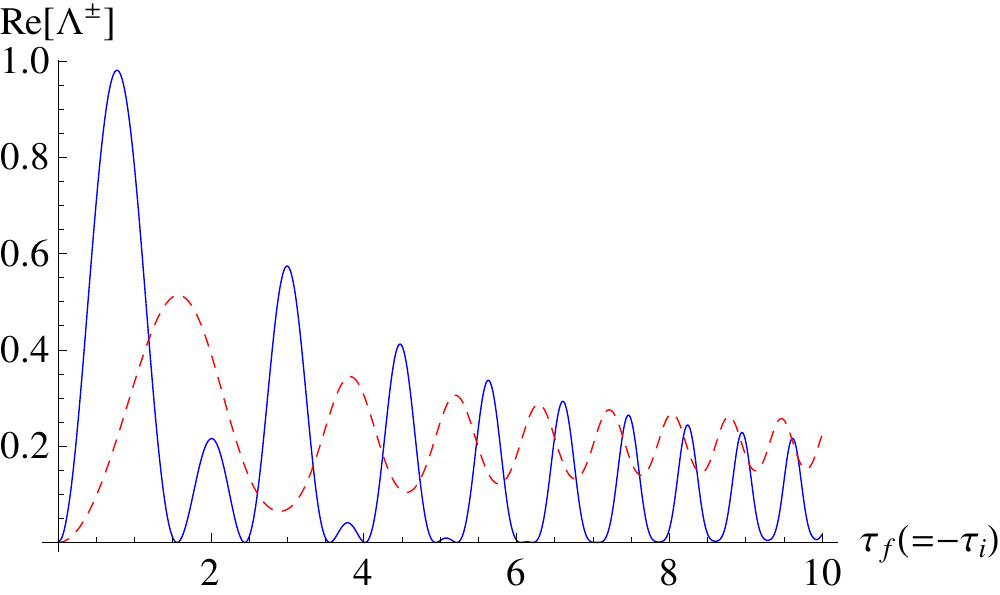}
} 
\subfigure{
\includegraphics[width=0.9\columnwidth]{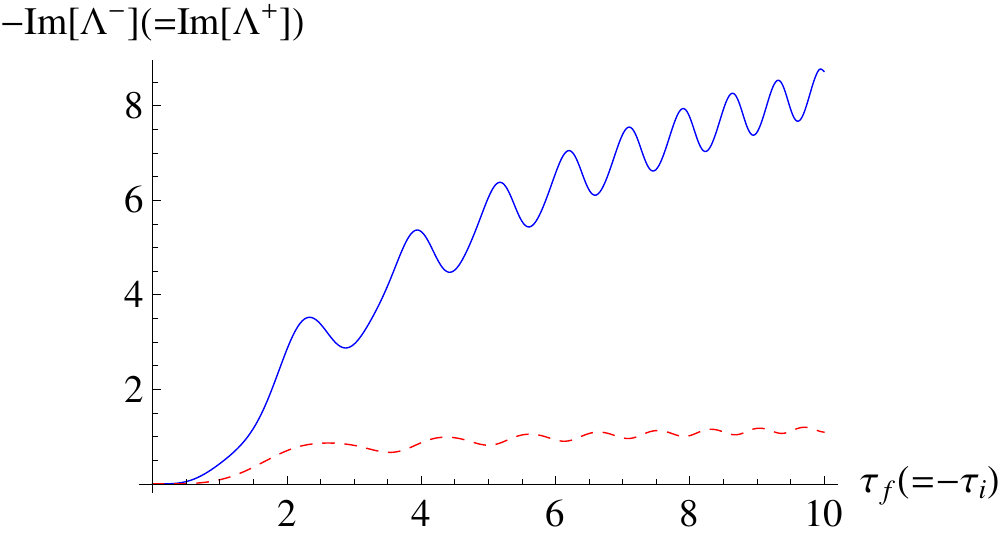}
}
\caption{Real part $P$ (a) and imaginary part $Q$ (b) of the function $%
\Lambda ^{+}(\protect \tau _{f})$ for a round sweep plotted \textit{vs}
half-sweep time $\protect \tau _{f}$. First part of the sweep, starting in
the $S$ state at $\protect \tau _{i}<\protect \tau <0$, stops in the
anticrossing point at $\protect \tau =0$ and runs immediately in the opposite
direction until $\protect \tau _{f}=|\protect \tau _{i}|$. Full (blue) lines $%
\protect \gamma =1$, dashed (red) lines $\protect \gamma =0.1$. }
\label{fig:Lambdasweetspot}
\end{figure}

To demonstrate the effect of the tunneling process near the anticrossing
point, in Fig.~\ref{fig:Lambdasweetspot} are plotted the data for a cycle
that begins in the $S$ state at $-\tau _{i}$, reaches the anticrossing at $%
\tau =0$, and then runs immediately back with the same speed until $\tau _{f}
$ with $\tau _{i}=\tau _{f}$ . Comparison of Figs.~3(a) and 4(a) for $\gamma
=1$ shows quite similar patterns of the oscillations of $P(\tau _{f})$ that
are more regular in Fig.~4(a). However, the patterns for $\gamma =0.1$ are
rather different demonstrating essential decrease in the spin transfer. The
magnitudes of $Q=\mathrm{Im}\{ \Lambda ^{+}\}$ are small in both cases, but
their $\tau _{f}$ dependences are rather different.

\subsection{Analytical theory of the probability oscillations}

\label{sec:anacycle}

We can explain the oscillations in the transition probability as a function
of the total duration of the cycle employing the analytical results in Sec. %
\ref{sec:LandauZener}. The \textit{forward} sweep from $-\tau _{i}$ to the
turning point $\tau _{m}$ gives rise to the singlet and triplet amplitudes
of Eq. (\ref{cSandcT+}). Assuming $\tau _{m}\gg 1$ and $\tau _{i}\gg 1$, and
employing (\ref{asympt1}) and (\ref{asympt4}), the singlet and triplet
amplitudes at the turning point $\tau _{m}$ 
are 
\begin{align}
\tilde{c}_{S}^{(fS)}& \approx e^{-\pi \gamma }e^{i\left( \tau _{i}^{2}-\tau
_{m}^{2}\right) /4}\tau _{i}^{i\gamma }\tau _{m}^{-i\gamma },  \label{cSm} \\
\tilde{c}_{T_{+}}^{(fS)}& \approx e^{-i3\pi /4}e^{-\pi \gamma /2}\frac{\sqrt{%
2\pi \gamma }}{\Gamma (1+i\gamma )}e^{i(\tau _{i}^{2}+\tau _{m}^{2})/4}\tau
_{i}^{i\gamma }\tau _{m}^{i\gamma }.  \label{cT+m}
\end{align}%
Here the superscripts indicate that we started in the singlet $S$ state and
carried out a forward linear sweep. The phase of the early time singlet
state is arbitrary and is omitted because it only modifies the overall phase
of the wave function and does not influence the final result for the
probability. The amplitudes of Eqs. (\ref{cSm}) and (\ref{cT+m}) are derived
under the assumption that $\tilde{c}_{S}=1$ and $\tilde{c}_{T_{+}}=0$ at
time $\tau =-\tau _{i}$.

Next, we consider the \textit{backward} sweep and include the contributions
from two channels passing through the $T_{+}$ and $S$ states at the turning
point. As discussed in Sec. \ref{sec:reverse}, the dynamical equations for
the amplitudes for the backward sweep $(\tilde{c}_{S}^{(b)},\tilde{c}%
_{T_{+}}^{(b)})$ differ from Eq. (\ref{Hst+transform}) by the interchange $%
\tilde{c}_{S}\longleftrightarrow \tilde{c}_{T_{+}}$. In order to make
contact with our results in Sec. \ref{sec:LandauZener}, we change the time $%
\tau \rightarrow \tau -2\tau _{m}$ for the backward sweep. Using the
interchange $\tilde{c}_{S}\longleftrightarrow \tilde{c}_{T_{+}}$, it follows
from Eq. (\ref{cT+}) that for the triplet $T_{+}$ channel the ratio of the
final and initial amplitudes along the backward sweep is $\tilde{c}%
_{S}^{(bT_{+})}/\tilde{c}_{T+}^{(fS)}=\sqrt{\gamma }e^{-i3\pi
/8}D_{-1-i\gamma }\left( -e^{i\pi /4}\tau _{f}\right) /\left[ e^{-i3\pi
/8}D_{i\gamma }\left( -e^{i3\pi /4}\tau _{m}\right) \right] ^{\ast }$, $%
T_{+} $ in the superscript of $\tilde{c}_{S}^{(bT_{+})}$ indicates the
channel. In the limit $\tau _{f}\gg 1$, Eqs. (\ref{asympt4}) and (\ref%
{asympt3}) imply that this ratio equals 
\begin{equation}
\tilde{c}_{S}^{(bT_{+})}/\tilde{c}_{T_{+}}^{(fS)}\approx e^{-i3\pi
/4}e^{-\pi \gamma /2}\frac{\sqrt{2\pi \gamma }}{\Gamma (1+i\gamma )}%
e^{i(\tau _{f}^{2}+\tau _{m}^{2})/4}\tau _{f}^{i\gamma }\tau _{m}^{i\gamma }.
\label{cSbacktriplet}
\end{equation}%
Similarly, by using the interchange $\tilde{c}_{S}\longleftrightarrow \tilde{%
c}_{T_{+}}$, it follows from Eq. (\ref{cT+tripinit}) that for the singlet $S$
channel the ratio of the final and initial amplitudes along the backward
sweep $\tilde{c}_{S}^{(bS)}/\tilde{c}_{S}^{(fS)}=D_{i\gamma }\left( e^{3i\pi
/4}\tau _{f}\right) /D_{i\gamma }\left( -e^{3i\pi /4}\tau _{m}\right) $. In
the limit $\tau _{f}\gg 1,$ Eqs. (\ref{asympt4}) and (\ref{asympt3}) imply
that the ratio of the singlet amplitudes after the backward sweep equals 
\begin{equation}
\tilde{c}_{S}^{(bS)}/\tilde{c}_{S}^{(fS)}\approx e^{-\pi \gamma }e^{i(\tau
_{f}^{2}-\tau _{m}^{2})/4}\tau _{f}^{i\gamma }\tau _{m}^{-i\gamma }\,.
\label{cSbacksinglet}
\end{equation}%
The singlet amplitude at the final time $\tau _{f}$ after the cycle of
duration $\left( \tau _{i}+\tau _{m}\right) +\left( \tau _{m}+\tau
_{f}\right) $ is a sum of the contributions coming from both channels, $%
\tilde{c}_{S}^{(tot)}=\tilde{c}_{S}^{(bT_{+})}+\tilde{c}_{S}^{(bS)}$.
Finally, 
\begin{eqnarray}
\tilde{c}_{S}^{(tot)} &\approx &e^{i\left( \tau _{f}^{2}+\tau _{i}^{2}-2\tau
_{m}^{2}\right) /4}\left( \frac{\tau _{f}\tau _{i}}{\tau _{m}^{2}}\right)
^{i\gamma }  \notag \\
&&\times \left[ \left( 1-P_{LZ}\right) +P_{LZ}e^{i\vartheta (\tau _{m})}%
\right] ,  \label{cStotalcyclic}
\end{eqnarray}%
where the Landau-Zener transition probability$P_{LZ}$ for a single-passage
is defined by Eq. (\ref{PLandauZener}) and the phase $\vartheta (\tau _{m})$
at the turning point $\tau _{m}$ is defined as 
\begin{equation}
e^{i\vartheta (\tau _{m})}=e^{i\tau _{m}^{2}}\tau _{m}^{4i\gamma }e^{-i\pi
/2}\Gamma (-i\gamma )/\Gamma (i\gamma ).  \label{phase}
\end{equation}
%

The dependence of the transition probability $P=1-\left \vert \tilde{c}%
_{S}(\tau _{f})\right \vert ^{2}$ on the position $\tau_m$ of the turning
point is 
\begin{equation}
P(\tau _{m}) =4P_{LZ}(1-P_{LZ})\sin ^{2}\vartheta (\tau _{m})/2,
\label{Pcyclic}
\end{equation}
%
where $\vartheta (\tau _{m})$ is the St\"{u}ckelberg phase.\cite%
{Stueckelberg,ShimshoniGefen,Shevchenko} It is acquired between the two
passages and includes both the adiabatic and non-adiabatic ($\gamma $%
-dependent) parts. From Eq.~(\ref{Pcyclic}) we can make several observations
that are consistent with the numerical data of Fig. \ref%
{fig:Lambdalinearbackandforth}. First, when $\tau _{f}\gg 1$ and $\tau
_{i}\gg 1$, $P$ does not depend on the initial and final times. The
transition probability only depends on the Landau-Zener probability $P_{LZ}$
of Eq. (\ref{PLandauZener}) and the turning point $\tau _{m}$. This means
that the oscillations of $P(\tau _{m})$ are a robust feature of a \textit{%
coherent} double passage across a Landau-Zener anticrossing. The transition
probability oscillates around the average value 
\begin{equation}
P_{av}=2P_{LZ}(1-P_{LZ}).  \label{Pav}
\end{equation}
%
For fast sweeps $P_{LZ}\ll 1$ so that $P$ oscillates between $0$ and $4P_{LZ}
$. For slow sweeps $P_{LZ}$ is close to $1$ and the probability oscillates
between $0$ and $4(1-P_{LZ})$. The maximum in the oscillation amplitudes is
achieved at $P_{LZ}=1/2$. When $P_{LZ}=1-e^{-2\pi \gamma }=1/2$ ($\gamma
\approx 0.11$), the transition probability $P$ oscillates between $0 $ and $1
$. The amplitudes of the oscillations are smaller for all other values of $%
\gamma $. This is exacly the behavoir we see in the numerical plots. One
more remarkable feature of Fig \ref{fig:Lambdalinearbackandforth}(a), that
all oscillations pass through $P=0$, is also reflected by Eq. (\ref{Pcyclic}%
).

Oscillatory patterns of $\gamma =0.1$ curves in Figs. 2(a) and 3(a) show
strikingly different behavior. In Fig.~2(a), the amplitude of oscillations
decreases with $\tau _{f}$, and $P$ gradually approaches its Landau-Zener
limit $P_{LZ}$. On the contrary, in Fig.~3(a) the oscillations, after some
transitional period, acquire a stationary amplitude. Eqs.~(\ref{Pcyclic})
and (\ref{Pav}) clarify the origin of this behavior typical of double
passages across the anticrossing.\cite{Stueckelberg,ShimshoniGefen,Shevchenko} Indeed, $%
P_{av}$ of Eq. (\ref{Pav}) is a Landau-Zener probability $P_{LZ}^{(2)}$ for
a double passage across the anticrossing that can be derived directly by the
above two-channel procedure with quantum amplitudes substituted by
probabilities, see Ref.~\onlinecite{LL:book65}. Therefore, suppression of
these long-time scale oscillations and approaching the double-passage
Landau-Zener limit $P_{LZ}^{(2)}$ are only achieved when the decoherence is
taken into account, and can allow measuring decoherence times.

In conclusion, prolonged oscillations of the electronic amplitudes $({\tilde{%
c}}_{S},{\tilde{c}}_{T_{+}})$ are a generic property of the coherent
electron dynamics during the single- and double-passages across the $S$-$%
T_{+}$ anticrossing. Their amplitudes and durations are controlled by the
Landau-Zener parameter $\gamma $ and by dephasing on longer time scales, and
the patterns are rather different for the single- and double-passages.

\section{Back action of nuclear spin dynamics on Overhauser fields}

\label{sec:SUP}


The Hamiltonian $\hat{H}$ of Eq. (\ref{Hfull}) describing the electron
states depends on the Overhauser fields created by the spatially dependent
nuclear spin configuration. The electrons experience the nuclear fields 
$v_n^\alpha$ and $\boldsymbol{\eta}_n$ of Eqs.\ (\ref{vn}) and (\ref{etaHat}%
), where the first represents the components of the effective difference
magnetic field in the dots, and the second represents the induced average
magnetic field. When going through the $S$-$T_{+}$ transition, the electrons
will experience a change of these nuclear Overhauser fields. It is a unique
property of Eq.~(\ref{totalchangeIz}) for the change in the total
longitudinal nuclear spin $\Delta I^{z}$ that it expresses a global property
of a double dot in terms of the parameters of the electronic Hamiltonian and
does not depend of a specific configuration of nuclear spins. For different
elements in the Hamiltonian $\hat{H}$, we calculate their mean-square values
as well as their variances.

The expression for the change of the total $z$ component of the nuclear spin
of ~Eq. (\ref{totalchangeIz2}) makes the role of $Q$ explicit due to the
mediation of spin-orbit coupling. With $v_{so}^{\pm }=0$, the total spin
transfer is protected by the momentum conservation law and $Q$ manifests
itself through shake-up processes in the nuclear spin reservoir respecting
the conservation of the total angular momentum. The electron dynamics
induces changes in the nuclear spin configuration that in turn induce
changes in the in the diagonal and off-diagonal elements of the electron
Hamiltonian (\ref{Hfull}). In what follows, we compute these changes.

\subsection{Changes in Overhauser fields}

\label{sec:VIIA}


Electrons experience an effective Zeeman splitting in the Overhauser field
of $\hat{\boldsymbol{\eta}}_j$ of Eq.~(\ref{etaHat}). The associated change
in the $z$-component of $\hat{\boldsymbol{\eta}}_j$, $\Delta{\hat \eta}%
_n^z=-A\sum_j\zeta_j\Delta{\hat I}_j^z$, is 
\begin{equation}
\Delta {\hat \eta}_n^z=\frac{A^2}{2v_\perp^2}\sum_j\rho_j\zeta_j(\Lambda^-v^-%
{\hat I}_j^++\Lambda^+v^+{\hat I}_j^-),  \label{SUp.0}
\end{equation}
In the multicycle regime, the field of Eq.~(\ref{SUp.0}) has been measured
by Petta et al.\cite{Petta2008} and by Foletti et al.\cite{Foletti:nphys09}
by the shift in the position of the $ST_+$ anticrossing. In contrast to $%
\Delta I^z$, the change $\Delta {\hat \eta}_n^z$ in the longitudinal field
depends on the detailed nuclear spin configuration and on the spatially
dependent electron-nuclear couplings $\rho_j$ of Eq. (\ref%
{singlettripletcoupling}) and $\zeta_j$ of Eq. (\ref{triplettripletcoupling}%
).

The singlet-triplet terms ${\hat v}_n^\pm$ and ${\hat v}_n^z$ in the
Hamiltonian $\hat H$ of Eq.~(\ref{Hfull}) are sums over all nuclear spins. $%
ST_0$ level splittings characterized by ${\hat v}_n^z$ were measured in Ref.~%
\onlinecite{Petta2005} and a number of follow-up papers, and $ST_+$
splittings described by ${\hat v}_n^\pm$ in Ref.~\onlinecite{Petta2010}. The
changes in these terms during a cycle are $\Delta {\hat v}%
^\alpha_n=A\sum_j\rho_j{\hat I}_j^\alpha$. By using Eq.~(\ref{deltaIj+-}),
we find changes in the components $\alpha=\pm$ that couple $S$ to $T_\pm$ 
\begin{eqnarray}
\Delta \hat{v}_n^{\pm } &=&\frac{A^2}{2v_{\bot }}\left[ \frac{v^{\pm }}{%
v_{\bot }}\Lambda ^{\pm }\sum_j\rho^2_{j}\hat{I}_{j}^{z}\pm i\Lambda
^{z}\sum_j\rho_j\zeta _{j}\hat{I}_{j}^{\pm }\right]  \notag \\
&&\mp iA\sum_j\rho_j\frac{\eta _{j(nZ)}}{\hbar }(T_{f}-T_{i})\hat{I}%
_{j}^{\pm } ,  \label{Dvntransverse}
\end{eqnarray}
%
and, by using Eq.~(\ref{deltaIz}), in the component $\alpha=z $ coupling $S$
to $T_0$ 
\begin{equation}
\Delta \hat{v}_{n}^{z}=-\frac{A^2}{2v_{\bot }^{2}}\left[ \Lambda
^{-}v^{-}\sum_j\rho _{j}^2\hat{I}_{j}^{+}+\Lambda ^{+}v^{+}\sum_j\rho^2 _{j}%
\hat{I}_{j}^{-}\right] .  \label{deltavz}
\end{equation}
%
We note that while ${\hat v}^z_n$ only produces a longitudinal Overhauser
field mixing $S$ and $T_0$, $\Delta{\hat v}_n^z$ includes operators ${\hat I}%
^\pm_j$ and therefore mixes $S$ and $T_+$ belonging to our $2\times2$
subspace.

In the next sub sections, mean values and variances of these operatores are
computed.

\subsection{Constraints and mean values}

\label{sec:VIIB}


While nuclear spins are distributed in the bath randomly, the magnetization
fluctuations $v_{n}^{\pm }$ controlling electron dynamics during the cycle
impose on their values the constraints 
\begin{equation}
A\sum_{j}\rho _{j}I_{j}^{\alpha }=v_{n}^{\alpha },  \label{constr}
\end{equation}
%
adding also a constraint related to $v_{n}^{z}$. To simplify calculations,
we consider below the nuclear spins $\mathbf{I}_{j}$ as random Gaussian
variables that are normalized, 
in the absent of constraints, as $\langle I_{j}^{\lambda }I_{j^{\prime
}}^{\lambda ^{\prime }}\rangle =\frac{1}{3}I_{j}(I_{j}+1)\delta _{jj^{\prime
}}\delta _{\lambda \lambda ^{\prime }}$, with $\lambda =(x,y,z)$. Then the
mean values of $I_{j}^{\lambda }$ are 
\begin{equation}
\langle I_{j}^{\lambda }\rangle =\frac{\int dI_{j}^{\lambda }I_{j}^{\lambda }%
\mathcal{P}(I_{j}^{\lambda })\prod_{j^{\prime }\neq j}\int dI_{j^{\prime
}}^{\lambda }\mathcal{P}(I_{j^{\prime }}^{\lambda })\delta (v_{n}^{\lambda
}-A\sum_{j^{\prime }}\rho _{j^{\prime }}I_{j^{\prime }})}{\prod_{j^{\prime
}}\int dI_{j^{\prime }}^{\lambda }\mathcal{P}(I_{j^{\prime }}^{\lambda
})\delta (v_{n}^{\lambda }-A\sum_{j^{\prime }}\rho _{j^{\prime
}}I_{j^{\prime }}^{\lambda })},  \label{Con1}
\end{equation}
%
where $\mathcal{P}(I_{j}^{\lambda })$ are Gaussian probabilities, $%
(v_{n}^{x},v_{n}^{y})$ are defined as $v_{n}^{\pm }=(v_{n}^{x}\pm v_{n}^{y})/%
\sqrt{2}$, and the denominator secures the normalization of the
probabilities under the constraints of Eq.~(\ref{constr}).

Using the integral representation for $\delta$-functions 
\begin{equation}
\delta(x)=\frac{1}{2\pi}\int_{-\infty}^\infty e^{i\omega x}d\omega,
\end{equation}
multiple Gaussian integrations of Eq.\ (\ref{Con1}) result in 
\begin{equation}
\langle I^\pm_j\rangle=\rho_jv_n^\pm/(AR_2),\, \, \langle
I^z_j\rangle=\rho_jv_n^z/(AR_2),  \label{Con2}
\end{equation}
%
where $R_n=\sum_j\rho_j^n$ are determined by the spatial dependence of the
electron-nuclear coupling constants. Substituting these expressions into
Eqs.~(\ref{SUp.0}) and (\ref{Dvntransverse}), we arrive at the corrections
to the nuclear field experienced by the electron spin during the sweep. 
\begin{equation}
\langle \Delta \eta^z_n\rangle=-\Delta I^z AR_3^\prime/R_2,  \label{Con3}
\end{equation}
where $R_3^\prime=\sum_j\rho^2_j\zeta_j$, and the Overhauser field mixing
its $S$ and $T_+$ components 
\begin{equation}
\langle \Delta v^z_n\rangle=-\Delta I^z AR_3/R_2,  \label{Con4}
\end{equation}
with $\Delta I^z$ of Eq.~(\ref{totalchangeIz}).

We see that both the changes in the longitudinal difference field $\Delta
v_n^z$ and the longitudinal average field $\Delta \eta_z$ are proportional
to the change in the total nuclear spin $\Delta I^z$. It follows from Eqs.~(%
\ref{singlettripletcoupling}) and (\ref{triplettripletcoupling}) that $\rho
_{j}$ typically have opposite signs in both dots while $\zeta _{j}>0$
everywhere, hence, $R_3^\prime>0$. Therefore, with $A>0$, the sign of $%
\langle \Delta \eta _{n}^{z}\rangle$ (the change in the mean Overhauser
field building in the double dot) is opposite to the sign of $\Delta I^z$,
in agreement with Eq.~(\ref{etaHat}). The sign of $\langle \Delta v^z_n\rangle
$ is defined by the sign $R_3$ that depends on the choice of electronic
basis functions (see Appendix B), therefore, it is not uniquely defined with
respect to $\Delta I^z$.

The magnitudes of $\Delta \eta_z$ and $\Delta v_n^z$ are of the order of $%
\Delta I^z An_0/N$ per cycle, i.e., about $\Delta I^z/\sqrt{N}$ of the mean
values of $\eta_z$ and $v_n^z$. For $v_{so}^\pm=0$, $\Delta I^z=-P$, hence, $%
\mid \Delta I^z\mid \leq1$. However, it is seen from Figs.~2(b) and 3(b) that $%
Q$ is an order of magnitude larger than $P$ when $\gamma \agt1$. Therefore,
when $v_{so}\neq 0$, the conditional expectation values $\langle \Delta \eta
_{n}^{z}\rangle $ and $\langle \Delta v_{n}^{z}\rangle $ should experience $Q
$-enhancement through the $Q$-enhancement of $\Delta I^{z}$, and $\eta_z$
and $v_n^z$ can change by about 1\% per cycle.

The mean values of the transverse components of $\mathbf{v}_{n}$, calculated
in a similar way from Eq.~(\ref{Dvntransverse}), are 
\begin{eqnarray}
\langle \Delta v_{n}^{\pm }\rangle &=&A\frac{v^{\pm }v_{n}^{z}}{2v_{\perp
}^{2}}\frac{R_{3}}{R_{2}}\Lambda ^{\pm }\pm iA\frac{v_{n}^{\pm }}{2v_{\perp }%
}\frac{R_{3}^{\prime }}{R_{2}}\Lambda ^{z}  \notag \\
&\mp &iv_{n}^{z}{\bar{\eta}}_{(nZ)}(T_{f}-T_{i})/\hbar  \label{Con5}
\end{eqnarray}%
%
where ${\bar{\eta}}_{(nZ)}$ is a mean value of $\eta _{j(nZ)}$ over all
nuclear species. Because different species are distributed randomly at the
scale of atomic spacings, they self-average in the linear approximation over 
$T_{f}-T_{i}$, and we accept that all of them have the same absolute values
of the angular momenta, $I_{j}=I$. While the first term is comparable in the
magnitude to Eq.~(\ref{Con4}), the two last term might be much larger
because they increase with the sweep duration. However, Eq.~(\ref{Con5})
includes changes both in the amplitude and the phase of $\Delta v_{n}^{\pm }$%
, and the latter might not be essential when solving Eq.~(\ref{Hst+transform}%
) that only depends on $v_{\perp }$. We come back to this term in Sec.~ \ref%
{sec:VIID}.

\subsection{$ST_+$-pulses induced interdot shake-ups}

\label{sec:VIIC}


Let us explain the importance of the variance in the spin production by
considering the total nuclear spins in the left and right dots. Average
values of different operators calculated in Sec.~\ref{sec:VIIB} were based
on the conditional mean values $\langle I^\alpha_j\rangle$ of nuclear spins $%
I_{j}^{\alpha }$ of the order of $N^{-1/2}$ that are small compared with
their root mean-square values. Therefore, calculating the mean-square values
of all operators and their variances is important for estimating the widths
of statistical distributions.

We begin with the differences in the spin polarizations of the left and
right dots, $L$ and $R$, that are critical for spin manipulation. While
division of a double dot into its left and right parts holds only when the
overlap integral is small enough, cf. Appendix B, the results are
instructive. Splitting Eq.~(\ref{vn}) into sums over $L$ and $R$, we define
partial sums 
\begin{equation}
v^\alpha_{nL(R)}=A\sum_{j\in L(R)}\rho_jI^\alpha_j.  \label{Shake1}
\end{equation}
Their sums are $v^\alpha_n$ and are a subject to constrains of Eq.~(\ref%
{constr}). However, their differences 
\begin{equation}
u^\alpha_n=v^\alpha_{nL}-v^\alpha_{nR}  \label{Shake2}
\end{equation}
are free of any constraints. Using Eq.~(\ref{deltaIz}), the change in the
left-right polarization difference is 
\begin{equation}
\Delta I^z_{LR}=-\frac{1}{2v^2_\perp}(\Lambda^-v^-u^+_n+\Lambda^+v^+u^-_n).
\label{Shake3}
\end{equation}
When averaged over an unpolarized spin reservoir, its mean value vanishes, $%
\langle \Delta I^z_{LR}\rangle=0$, and the mean-square value equals 
\begin{equation}
\langle(\Delta I^z_{LR})^2\rangle=\frac{A^2n_0}{6v_\perp^2}I(I+1)\mid
\Lambda \mid^2\int \rho^2(\mathbf{R})d^3\mathbf{R},  \label{Shake4}
\end{equation}
with $\rho(\mathbf{R})$ of Eq.~(\ref{singlettripletcoupling}) and 
\begin{equation}
\mid \Lambda \mid^2=P^2+Q^2.  \label{Shake5}
\end{equation}
A simple estimate of the right hand side of Eq.~(\ref{Shake4}) results in $%
\mid \Lambda \mid^2$. Therefore, the asymmetry of spin pumping of the left
and right dots is $Q$-enhanced whenever $Q\gg P$, in particular, when $%
v_{so}=0$ and $P\leq1$. We attribute this enhancement to \textit{shake-up
processes} resulting in multiple spin flips per each ``pure" injected
nuclear spin. These processes are random, and it is not clear for now how
they influence inhomogeneous spin distributions.\cite%
{Ramon:prb07,Gullans:prl10}

The detailed spatial patterns of spin generation at long time scales are a
subtle subject and are related to the spatial variation of the
electron-nuclear couplings $\rho (\mathbf{R}_{j})$ and $\zeta (\mathbf{R}%
_{j})$ calculated in Appendix \ref{sec:spatial}. With mean values of $%
I_{j}^{\pm }$ of Eq.~(\ref{Con2}), spatial distribution of $\Delta I_{j}^{z}$
is related to $\Delta I^{z}$ as $\Delta I_{j}^{\pm }=(\rho
_{j}^{2}/R^{2})\Delta I^{z}$. The left-right asymmetry in $\rho _{j}^{2}$
originates either from the geometric asymmetry of the double dot\cite%
{Gullans:prl10} or from the $L$-$R$-overlap of the electron density, cf.
Appendix B, and produces a regular difference in the $I^{z} $ generation
rate. While the results depend on the specific distribution of nuclear spins
and the $S$-$T_{0}$ mixing,\cite{Stopa:prb10} the mechanism of $Q $%
-enhancement is quite general whenever $\gamma \agt1$.

\subsection{Mean-square values and variances}

\label{sec:VIID}


Mean values of Sec.~\ref{sec:VIIB} were evaluated over an unpolarized
nuclear spin bath and estimate the mean rates of the change of the different
parameters. However, the estimate of the shake-up rate of Sec.~\ref{sec:VIIC}
demonstrates that calculating variances of these random variables can
provide additional, and sometimes even more valuable, information about the
magnitudes of the expected changes during a cycle. The conditional
probability distributions are so wide that the mean value is not very
representative. In this section, we evaluate variances of the basic nuclear
fields. 

We begin with calculating the mean-square values. Because all nuclear fields
of Eqs.~(\ref{SUp.0}) - (\ref{deltavz}) are linear in the momenta $I^\alpha_j
$, mean values of the quadratic forms in them include integrals that differ
from Eq.~(\ref{Con1}) by substituting $I^\lambda_j$ either by $%
(I^\lambda_j)^2$ or by $I^\lambda_{j}I^\lambda_{j^\prime}$ with $j\neq
j^\prime$. While the latter terms are smaller in the parameter $1/N\ll1$,
they have a higher statistical weight. Summing all terms, one arrives at
length expressions for $\langle(\Delta \eta_n^z)^2\rangle$ and $%
\langle(\Delta v_n^z)^2\rangle$ that we do not present here. Instead, using
the mean values of Eqs.~(\ref{Con3}) and (\ref{Con4}), we present the
variances defined as $\mathrm{Var}\{ \mathrm{\xi}\}=\langle
\xi^2\rangle-\langle \xi \rangle^2$ 
\begin{equation}
\mathrm{Var}\left \{ \Delta{\eta}^z_n\right \}=\mid \Lambda \mid^2\frac{A^4}{%
6v_\perp^2}I(I+1) [R_4^\prime-(R_3^\prime)^2/R_2],  \label{MS.1}
\end{equation}
where $R_4^\prime=\sum_j(\rho_j\zeta_j)^2$, and 
\begin{equation}
\mathrm{Var}\left \{ \Delta{v}_n^z\right \}=\mid \Lambda \mid^2\frac{A^4}{%
6v_\perp^2}I(I+1) [R_4-(R_3)^2/R_2].  \label{MS.2}
\end{equation}

Comparison with Eqs.~(\ref{Con3}) and (\ref{Con4}) shows $Q$-enhancement
even when $v_{so}=0$ (hence, when $\Delta I=-P)$, the effect that manifested
itself already in Eq.~(\ref{Shake4}). 
This means that the nuclear spins with $I_j^z$ away from the mean
conditional expectation values $\langle I_j^z\rangle$ respond to the sweeps
stronger than the spins with $I_j^z=\langle I_j^z\rangle$. Also, this
enhanced sensitivity is due to the spatial distribution of $\rho_j$ and $%
\zeta_j$ because with $\rho_j$=const and $\zeta_j$=const the brackets in
Eqs.~(\ref{MS.1}) and (\ref{MS.2}) vanish. By the order of magnitude, both
quantities experience changes of about $\Lambda An_0/N$ per cycle; with $%
\Lambda \approx10$ and $N\approx10^6$, this suggests changes about 1\% per
cycle. In other words, around 10 spins interchange their directions during
one passage.

Calculating $\langle \Delta (v_{n}^{+}v_{n}^{-})\rangle $ results in a
simple equation 
\begin{equation}
\langle \Delta (v_{n}^{+}v_{n}^{-})\rangle =-I^{z}Av_{n}^{z}R_{3}/R_{2}
\label{MS.3}
\end{equation}%
%
because the contributions of the two last terms of Eq. (\ref{Dvntransverse})
cancel. In absence of spin-orbit coupling, this immediately suggests $%
\langle \Delta (v_{\perp }^{2})\rangle =PAv_{n}^{z}R_{3}/R_{2}$. Under these
conditions, large terms in Eq. (\ref{Dvntransverse}) reflect only the change
in the phase of $v_{n}^{\pm }$ that does not influence dynamical equations (%
\ref{Hst+transform}), and the relative change in $(v_{n}^{\perp
})^{2}=v_{n}^{+}v_{n}^{-}$ is only about $N^{-1/2}$.

However, in presence of spin-orbit coupling the dynamics of spin amplitudes $%
({\tilde{c}}_{S},{\tilde{c}}_{T})$ is controlled by $v^{\pm }$ rather then $%
v_{n}^{\pm }$. Mean value of $\Delta (v_{\perp }^{2})$, calculated by using
Eqs.~(\ref{Dvntransverse}) and (\ref{Con2}), is 
\begin{eqnarray}
&&\langle \Delta (v^{+}v^{-})\rangle =\langle \Delta
(v_{n}^{+}v_{n}^{-})\rangle  \notag  \label{MS.4} \\
&+&\frac{Av_{n}^{z}}{2v_{\perp }^{2}}\frac{R_{3}}{R_{2}}[\Lambda
^{-}v_{n}^{-}v_{so}^{+}+\Lambda ^{+}v_{n}^{+}v_{so}^{-}]  \notag \\
&+&i(v_{n}^{-}v_{so}^{+}-v_{n}^{+}v_{so}^{-})\left[ \frac{{\bar{\eta}}_{nB}}{%
\hbar }(T_{f}-T_{i})-\frac{A}{2v_{\perp }}\frac{R_{3}^{\prime }}{R_{2}}%
\Lambda ^{z}\right] ,  \notag \\
&&
\end{eqnarray}%
%
where first term is defined by Eq.~(\ref{MS.3}). Physically, second and
third terms in Eq.~(\ref{MS.4}) take into account the 
angle between $v_{n}^{+}$ and $v_{so}^{+}$ in the complex plane, and are
proportional to the product $v_{\perp }v_{n}^{\perp }$. With $v_{\perp }\sim
v_{n}^{\perp }$, relative corrections coming from the second term are of the
order $\Lambda /\sqrt{N}$ per cycle. The third term is usually much larger
because it increases linearly with the pulse duration $\Delta T=T_{f}-T_{i}$%
. It includes two contributions of which first is due to the Zeeman
precession of nuclei and second due to the Knight field and is proportional
to the integral of $\mid {\tilde{c}}_{T+}\mid ^{2}$. While the magnitude of
the second contribution depends on the shape of the pulse, the ratio of
these terms is roughly ${\bar{\eta}}_{(nB)}/(An_{0}/N)$ and they become
comparable at $B\sim 1$ mT. This indicates that first contribution to the
third term usually dominates. With $v_{\perp }\sim v_{so}^{\perp }$ and ${%
\bar{\eta}}_{(nB)}\approx 10$ mT, the Zeeman term results in $\langle \Delta
(v^{+}v^{-})\rangle \sim 0.1\langle v^{+}v^{-}\rangle $ for a 0.1$\mu $s
linear sweep. This is much larger than the correction to the same quantity
estimated in Eq.~(\ref{MS.3}) and to $\langle (\Delta v_{n}^{z})^{2}\rangle $
having the same scale. The effect in InAs should be much larger than in GaAs
because of the stronger spin-orbit coupling.

The above estimates indicate that, because of the terms in Eq. (\ref%
{Dvntransverse}) linear in the pulse duration, spin-orbit corrections to
transverse matrix elements are essentially larger than the corrections to
the longitudinal ones.

In Eq.~(\ref{MS.4}), Zeeman precession of nuclei manifests itself in $%
\langle(v^+v^-)\rangle$ only through spin-orbit coupling. The effect is much
stronger when estimated through the variance of $v^+v^-$, and we estimate it
for $v_{so}=0$ when $v^\pm=v_n^\pm$. Disregarding two first terms in Eq.~(%
\ref{Dvntransverse}), the calculations similar to those performed above when
deriving Eqs.~(\ref{MS.1}) and (\ref{MS.2}) result in 
\begin{eqnarray}
\mathrm{Var}\{ \Delta(v_n^+v_n^-)\}&\approx&\frac{I(I+1)}{6}(\overline{%
\eta^2_{(nB)}}-{\bar \eta_{(nB)}}^2)  \notag \\
&\times&(v_n^+v_n^-)A^2R_2[(T_f-T_i)/\hbar]^2,  \label{MS.5}
\end{eqnarray}
where $\overline{\eta^2_{(nB)}}$ is the mean-square value of $\eta_{j(nB)}$.
It follows from Eq.~(\ref{MS.5}), the dominating mechanism of changing $%
v_n^\perp$ is the nuclear spin precession with a characteristic time of
about a microsecond at $B\sim10$ mT. It is about two to three orders of
magnitude shorter than the corresponding time for $v_n^z$ estimated above.

It is also instructive to compare this estimate with a much longer time for $%
v_n^\perp$ following from Eq.~(\ref{MS.3}). The latter estimate was found
with the nuclear configuration of Eq.~(\ref{Con2}) that reflects the
mean-values of nuclear spins under the constraints of Eq.~(\ref{constr}). In
a narrow region of the phase space around these mean values dynamics of $%
v_n^\perp$ is strongly suppressed. The estimate of Eq.~(\ref{MS.5}) is much
more representative because it represents the entire phase space compatible
with the constraints of Eq.~(\ref{constr}). A similar type of the behaviour
of $v_n^z$ was discussed above as applied to Eq.~(\ref{MS.2}).

\section{Conclusions}

\label{sec:conclusions}

We have studied the dynamics of the electron and nuclear spins near $ST_{+}$
avoided crossings in double quantum dots. While adopting the traditional
approach based on the hierarchy of time scales, with a slow nuclear and fast
electron dynamics, we employed a quantum description of the electron spin
and coherent dynamics of nuclear spins, and investigated the time-resolved
patterns of single and double Landau-Zener passages through the anticrossing
point. They are described by two complex conjugate functions $\Lambda ^{\pm
} $ depending on the initial and finite times $(T_{i},T_{f})$ and the
trajectory of the sweep, with $\Lambda ^{-}$ proportional to the integral of
the product ${\tilde{c}}_{S}^{\ast }(t){\tilde{c}}_{T}(t)$ of the complex
amplitudes of the $S$ and $T_{+}$ states. Their real parts $P=\mathrm{Re}\{
\Lambda ^{\pm }\}$ are proportional to the $S$-$T_{+}$-transition
probability and for one-side sweeps oscillate at small time scales when the
system is close to the anticrossing and saturate at long time scales. For
linear sweeps, we find the singlet and triplet amplitudes in terms of Weber $%
D$-functions (parabolic cylinder functions); the long-time asymptotic limit
of $P$ equals the Landau-Zener probability $P_{LZ}=1-e^{-2\pi \gamma }$. For
round trips, the system also experiences long-term St\"{u}ckelberg
oscillations. The first sharp oscillations might be utilized for ultrafast
electron spin operation, while the decay of the oscillations can provide
information about dephasing rates. It is important that the imaginary part $%
Q=\mathrm{Im}\{ \Lambda ^{+}\} $ that acquires contributions from the
electronic states at a wide time scale and accumulates with time (it
diverges logarithmically for linear sweeps) has a profound effect on the
dynamics of the nuclear spins. When the Landau-Zener parameter $\gamma
\gtrsim 1$, $Q$ is typically one order of magnitude larger than $P$.
Therefore, in presence of the spin-orbit coupling violating the angular
momentum conservation, $Q$ may become the major factor controlling the
angular momentum transfer to nuclei. In particular, this mechanism is
efficient for excursions including a 
stay near the anticrossing point. Generically, $\Lambda =(P^{2}+Q^{2})^{1/2}$
controls the shake-up processes that exchange angular momentum between the
left and right dots. With $Q\gg P$, it is $Q$ that plays a dominating role
in these angular-momentum exchange processes. Because the mechanism that
plagues many experimental efforts of building considerable polarization
gradients remains unknown, it is a challenging question whether and how the
shake-up processes contribute to it; unfortunately, only a theory including
multiple passages can resolve it. We also estimated changes in the
Overhauser fields during a single cycle and concluded that the transverse
components are more volatile than the longitudinal ones.

We are grateful to B. I. Halperin, C. M. Marcus, I. Neder, and M. Rudner for
stimulating discussions and comments on the manuscript. E. I. R. was funded
by IARPA through the Army Research Office, by NSF under Grant No.
DMR-0908070, and in part by Rutherford Professorship (Loughborough, UK).

\appendix

\section{Spin Operator}

\label{sec:spinoperator}

We use the following convention for the spin-1 operator $\mathbf{S}$%
\begin{eqnarray}
S_{x} &=&\frac{1}{\sqrt{2}}\left( 
\begin{array}{ccc}
0 & 1 & 0 \\ 
1 & 0 & 1 \\ 
0 & 1 & 0%
\end{array}%
\right) , \\
S_{y} &=&\frac{1}{\sqrt{2}}\left( 
\begin{array}{ccc}
0 & -i & 0 \\ 
i & 0 & -i \\ 
0 & i & 0%
\end{array}%
\right) , \\
S_{z} &=&\left( 
\begin{array}{ccc}
1 & 0 & 0 \\ 
0 & 0 & 0 \\ 
0 & 0 & -1%
\end{array}%
\right) .
\end{eqnarray}%
These operators satisfy the commutation relations$\left[ \hat{S}_{i},\hat{S}%
_{j}\right] =i\epsilon _{ijk}\hat{S}_{k}$, where $\  \epsilon _{ijk}$ is the
Levi-Civita tensor, as well as$\  \hat{S}_{x}^{2}+\hat{S}_{y}^{2}+\hat{S}%
_{z}^{2}=2$.

\section{Simple Model}

\label{sec:spatial}

The singlet part of the spin wave function is 
\begin{equation}
\chi _{S}(1,2)=\frac{1}{\sqrt{2}}\left( |\uparrow _{1}\rangle |\downarrow
_{2}\rangle -|\downarrow _{1}\rangle |\uparrow _{2}\rangle \, \right)
\label{T0spin}
\end{equation}%
%
and the three triplet components of the spin wave function are 
\begin{subequations}
\begin{align}
\, \chi _{T_{+}}(1,2)& =|\uparrow _{1}\rangle |\uparrow _{2}\rangle , \\
\chi _{T_{0}}(1,2)& =\frac{1}{\sqrt{2}}\left( |\uparrow _{1}\rangle
|\downarrow _{2}\rangle +|\downarrow _{1}\rangle |\uparrow _{2}\rangle
\right) , \\
\chi _{T_{-}}(1,2)& =|\downarrow _{1}\rangle |\downarrow _{2}\rangle .
\end{align}

We will in this section discuss the spatial dependence of the hyperfine
coupling constants $\rho _{j}$ of Eq. (\ref{singlettripletcoupling})\ and $%
\zeta _{j}$ of Eq. (\ref{triplettripletcoupling}). In a simple model, the
electron wave functions near the $S$-$T_{+}$ anticrossing are 
\end{subequations}
\begin{eqnarray}
\psi _{S}(1,2) &=&\cos \nu ~\psi _{R}(1)\psi _{R}(2)  \notag \\
&+&\frac{\sin \nu }{\sqrt{2}}[\psi _{L}(1)\psi _{R}(2)+\psi _{L}(2)\psi
_{R}(1)],  \label{singletrealization}
\end{eqnarray}%
\begin{equation}
\psi _{T}(1,2)=\frac{1}{\sqrt{2}}[\psi _{L}(1)\psi _{R}(2)-\psi _{L}(2)\psi
_{R}(1)],  \label{tripletrealization}
\end{equation}%
where $L$ denotes the left and $R$ the right dot, and the angle $\nu $
depend on the Zeeman energy $\eta _{Z}$. The normalization coefficients in (%
\ref{singletrealization}) and (\ref{tripletrealization}) are exact under the
assumption that the functions $\psi _{L}$ and $\psi _{R}$ are
orthonormalized.

Let us illustrate the spatial dependence of the electron-nuclear coupling
constants $\rho $ of Eq. (\ref{singlettripletcoupling}) and $\zeta $ of Eq. (%
\ref{triplettripletcoupling}) for a simple model of a quantum dot. We assume
the electrons are in the lowest orbital harmonic oscillator state. The
Cartesian coordinates, the wave function is $\psi (x,y)=\exp \left[
-(x^{2}+y^{2})/l^{2}\right] /(l\sqrt{2/\pi })$, where $l$ is the size of
each quantum dot. We have two quantum dots that are separated at a distance $%
d$, one at $x=-d/2$ and $y=0$ and the other at $x=d/2$ and $y=0$. We form an
orthonormal basis set based on the functions $\psi (x-d/2,y)$ and $\psi
(x+d/2,y)$. In this basis, we compute $\rho (x,y)$ and $\zeta (x,y)$.

\begin{figure}[tbph]
\centering
\subfigure[]{
\includegraphics[width=0.9\columnwidth]{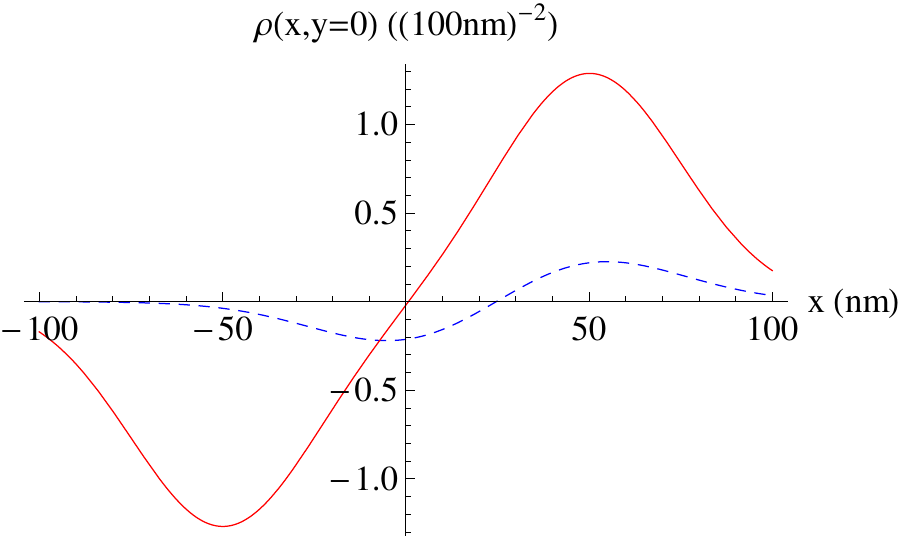}
\label{fig:elnucrho}
} 
\subfigure[]{
\includegraphics[width=0.9\columnwidth]{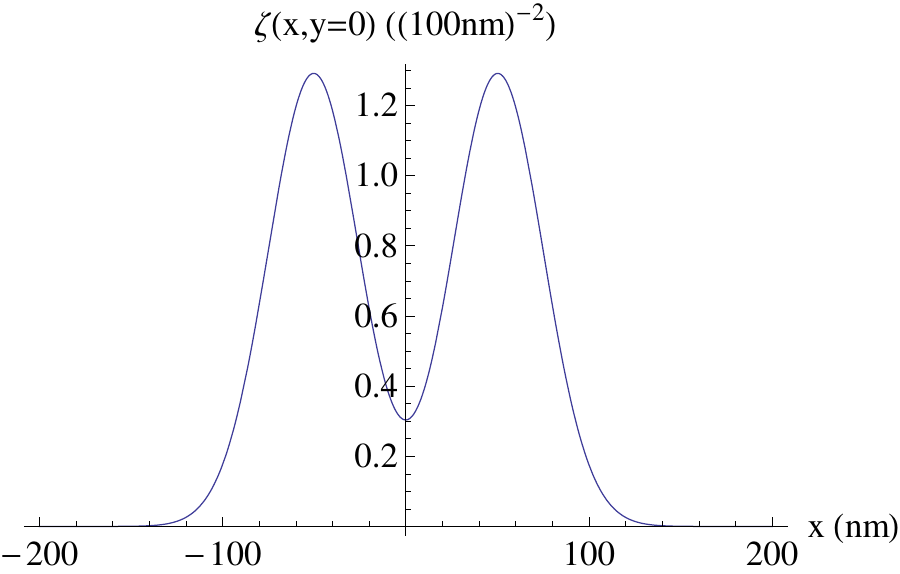}
\label{fig:elnucxi}
}
\caption{The spatial variation of the electron-nuclear couplings $\protect%
\rho (x,y=0)$ (a) and $\protect \zeta (x,y=0)$ (b). In (a), the red (full)
curve is for $\protect \nu =\protect \pi /2-0.1$ and the blue (dashed) curve
is for $\protect \nu =0.1$. The size of the dots is $l=50$ nm and the
separation between the dots is $d=100$ nm. The overlap integral between the
left and the right oscillator wave function is $0.1$. It is the most
striking feature that the overlap between the wave functions induces
asymmetry of the $\protect \rho(x,y)$ even in geometrically symmetric double
dots. The asymmetry reaches its maximum when the system is close to the $%
(0,2)$ state.}
\label{fig:elnuc}
\end{figure}

We plot in Fig.\  \ref{fig:elnuc} the electron-nuclear couplings $\rho (x,y)$
and $\zeta (x,y)$ for 
$y=0$ as a function of $x$ when $\nu =0.1$ and 
$\nu =\pi /2-0.1$. The spatial distribution of the singlet-triplet coupling $%
\rho (x,y)$ depends on the angle $\nu $. When $\nu $ is close to $\pi /2$,
there is a nearly equal probability for electrons to be located in the left
and right dot for both the singlet and triplet states. Then the
singlet-triplet coupling $\rho (x,y)$ is nearly antisymmetric around $x=0$, $%
\rho (x,y) \approx -\rho (-x,y) $ [the sign of $\rho (x,y)$ depends on the
sign choice in Eq. (\ref{tripletrealization})]. When $\nu $ is small, the
electrons are in the singlet state $(0,2) $ in the right dot, $\ $so that $%
\rho (x,y)$ passes through zero inside the right dot (for $x>0$). Therefore,
even for two symmetrically shaped dots, the $S$-$T_{+}$ electron-nuclear
coupling can become asymmetric because of the overlap of the left and the
right dot wave functions. The asymmetry depends on $\nu $ controlled by the
external magnetic field. 

The triplet-triplet electron-nuclear coupling $\zeta (x,y)$ does not depend
on $\nu $ and is a symmetric function of $x$ for the two symmetric quantum
dots.

\section{Two identities for the parabolic cylinder $D$-functions}

\label{sec:identities}


Using the solution of Eq.~(\ref{cSandcT+}) for ${\tilde c}_{T_+}(\tau)$ and $%
{\tilde c}_S(\tau)$ and the normalization condition $|{\tilde c}_S(\tau)|^2+|%
{\tilde c}_{T_+}(\tau)|^2=1$, we arrive at an identity 
\begin{equation}
\gamma \mid D_{-1-i\gamma}(-e^{i\pi/4}\tau)\mid^2+\mid
D_{i\gamma}(e^{i3\pi/3}\tau)\mid^2=e^{\pi \gamma/2}  \label{ID.1}
\end{equation}
relating absolute values of two $D$-functions at arbitrary real values of $%
\tau$ and $\gamma$.

Next, it follows from Eq.~(\ref{LZSchrodinger}) that 
\begin{equation}
\partial_\tau \left(\mid{\tilde c}_S\mid^2-\mid{\tilde c}_{T_+}\mid^2%
\right)=-2i\sqrt{\gamma}\left({\tilde c}_S^*{\tilde c}_{T_+}-{\tilde c}_S{%
\tilde c}_{T_+}^*\right).  \label{ID.2}
\end{equation}
Integrating it over $\tau$ and using Eqs.~(\ref{cSandcT+}) and (\ref%
{PLandauZener}), we find 
\begin{eqnarray}
&&\int_{-\infty}^\infty d\tau ~\mathrm{Im} \{e^{-i3\pi/4}D_{-1-i%
\gamma}(-e^{i\pi/4} \tau)D_{i\gamma}(e^{i3\pi/4}\tau)\}  \notag \\
&=&-\frac{\sinh \pi \gamma}{\gamma}e^{-\pi \gamma/2}.  \label{ID.3}
\end{eqnarray}
The integral of the real part of the integrand diverges.

While we could not find these identities for $D_n(z)$ functions with complex
(imaginary) indeces $n$ and the arguments directed along diagonals in the
complex $z$ planes in any of mathematical sources, we checked them
numerically.


\begin{thebibliography}{99}
\bibitem{Zenercomment} In Zener's paper \cite{Zener:prs32}, the aymptotics
of Ref. \cite{WhittakerWatson:book45} have been used. Fortunately, this did
not influence his final result for the Landau-Zener transition probability.
\end{thebibliography}
\end{document}